\documentclass[fleqn,usenatbib]{mnras}
\usepackage{multirow}

\usepackage[T1]{fontenc}
\usepackage{ae,aecompl}

\usepackage[dvips]{graphicx}	
\usepackage{amsmath}	
\usepackage{amssymb}	

\title[ Gas Metallicity gradients of Discs ]{The gas metallicity gradient and the star formation activity of disc galaxies}
\author[ Tissera et al.]{Patricia B. Tissera$^{1,2}$\thanks{E-mail:
patricia.tissera@unab.cl}, Susana E. Pedrosa$^{3}$, Emanuel Sillero$^{4}$ \& Jose M.Vilchez$^{5}$ \\
$^{1}$Departamento de Ciencias Fisicas, Universidad Andres Bello,
Av. Republica 220, Santiago, Chile. \\
$^{2}$Millennium Institute of Astrophysics, Av. Republica 220, Santiago, Chile.\\
$^{3}$Instituto de Astronom\'ia y F\'isica del Espacio (IAFE,
CONICET-UBA), C.C. 67 Suc. 28, C1428ZAA Ciudad de Buenos Aires, Argentina.\\
$^{4}$Instituto de Astronomia Te\'orica y Experimental (CONICET-UNC), Laprida 925, Cordoba, Argentina.\\
$^{5}$Instituto de Astrof\'isica de Andaluc\'ia (CSIC), Glorieta de la Astronom\'ia s/n, E-18008 Granada, Spain.\\
}

\date{Accepted XXX. Received YYY; in original form ZZZ}

\pubyear{2015}

\begin{document}
\label{firstpage}
\pagerange{\pageref{firstpage}--\pageref{lastpage}}
\maketitle

\begin{abstract}
 We study  oxygen abundance profiles
of the  gaseous disc components in  simulated galaxies  in a hierarchical universe.  We analyse the disc metallicity gradients in relation to the stellar masses and star formation rates of the simulated galaxies.
We find  a trend for galaxies with low stellar masses to have steeper metallicity gradients than  galaxies with 
 high  stellar masses at $z\sim 0$.
 We also detect that the gas-phase metallicity slopes and the
specific star formation rate (sSFR) of our simulated  disc galaxies  are consistent with recently reported observations  at $z\sim 0$. 
Simulated galaxies with high stellar masses 
reproduce the observed relationship  at all
analysed redshifts and have  an increasing contribution of discs with positive metallicity slopes with increasing redshift.
 Simulated galaxies with low stellar masses a have larger fraction of  negative  metallicity gradients with increasing redshift. 
 Simulated galaxies with positive or very negative  metallicity slopes exhibit disturbed morphologies and/or have a close neighbour. We analyse the evolution of the  slope of the oxygen profile and sSFR 
for a gas-rich galaxy-galaxy encounter, finding that this kind of events could generate either positive and negative  gas-phase oxygen profiles depending on  their state of evolution.
Our results support claims that the determination of reliable metallicity gradients  as a function of redshift is a key piece of information  to understand galaxy formation and set
constrains on the subgrid physics.

\end{abstract}

\section{Introduction}

Understanding the evolution of the chemical content of the baryons is  a crucial piece in the galaxy formation puzzle. Chemical elements are synthesized in the stellar interiors  and injected into the  interstellar medium (ISM) by Supernovae (SNe) and stellar winds. Part of the enriched material is locked up into stars
while the rest is mixed  with the existing ISM or expelled by galactic winds. Chemical abundances are also affected  by the combined action of dynamical processes such as   gas inflows and outflows, secular evolution, mergers and interactions, gas stripping. In  the current cosmological paradigm, these processes take place as galaxies are assembled in a non-linear way.
As a consequence, they   imprint  chemical patterns  in the stellar populations and the ISM,  which
  store valuable information  for galaxy formation studies \citep[e.g.][]{freeman} and to constrain the modelling of physical processes as  shown by analytical \citep[e.g.][]{matteucci1986,chiap1997,molla1995,molla1997,chiappini2001}, semi-analytical \citep[e.g.][]{cora2006,delucia2014} and numerical \citep[e.g.][]{mosconi2001,koba2007,wier2009} codes which consider the chemical evolution. In the last decades, hydrodynamical cosmological simulations have improved by incorporating more sophisticated schemes  for  relevant physical processes \citep[e.g.][]{schaye2015}. However, the baryonic physics is still mostly described by subgrid modelling which
need the adjustment of free parameters to reproduce observational constrains. Chemical abundances can be powerful tools to confront these
models with observations, providing further insights in the process of galaxy assembly within a cosmological context \citep[e.g.][]{tissera2012,gibson2013,aumer2013}.

 In the Local Universe,  metallicity gradients in disc  galaxies are found to be negative, consistently with an inside-out formation \citep[e.g.][]{sanchez2013Califa, sanchez2014}. 
Positive gradients have been measured in  interacting galaxies \citep[e.g.][]{rupke2010,rich2012,sanchez2014} which could be ascribed to  the triggering of low-metallicity
gas inflows during these events.  
This effect produced by massive metal-poor gas inflows has been also observed in the so-called "Green Peas" (GP) galaxies, which show extremely high SFR for their stellar mass mass together with a very low O/H abundance, as a consequence of the accreted metal-poor gas  \citep{amorin2010}. However, when  the relation of N/O versus the stellar mass is analysed, GPs are placed at the locus expected for their masses according to the general N/O versus mass relation. This fact reinforces the massive inflow scenario since the N/O ratio is insensitive (in first order) to gas inflow  \citep[e.g.][]{amorin2012}.
 At high redshift, the situation
seems to be more complex with disc galaxies reported to exhibit almost flat/positive \citep[e.g.][]{queyrel2012} and  negative  \citep[e.g.][]{swinbank2012,jones2013,yuan2011} abundance
gradients. It is important to bare in mind that the determination of abundance gradients can be very sensitive to the metallicity indicator and spectral resolution \citep[e.g.][]{kewley2008,marino2013,mast2014}.

Recently, \citet{stott2014} combined the metallicity gradient with the sSFR  of star-forming galaxies,  claiming the possible existence of a  correlation between them so that 
positive gradients would  be associated to starburst galaxies (high sSFR galaxies). 
They speculated that mergers, interactions or  efficient gas accretion into the central regions could be responsible of this correlation as these processes could contribute with low-metallicity gas inflows.
Testing 
 the existence of this relationship  and identifying the main processes responsible of its origin are of large interest to understand 
galaxy formation within the current cosmological model. 

Regarding the effects of mergers and interactions, observations of the central metallicity in galaxy pairs  indicate a trend for these systems to have lower abundances with respect to galaxies in isolation, on average \citep{kewley2006,ellison2008,dansac2008,dimat2009,kewley2010}. Several numerical simulations showed that mergers and interactions are efficient mechanisms at triggering inward gas inflows which can  feed  new star formation activity in galaxies \citep[e.g.][]{bh96,tissera2000}. If disc galaxies have negative metallicity gradients, then, these gas inflows can dilute the central metallicities. By using cosmological hydrodynamical simulations, \citet{perez2006} found a clear trend for galaxies in close pairs to exhibit lower central metallicity compared to galaxies in isolation, in agreement with observations. \citet{rupke2010}  reported   the flattening of the metallicity profiles in interacting galaxies using N-body galaxy-galaxy  mergers. More recently, \citet{perez2011}  studied in detail the evolution of the metallicity gradients during major mergers by using hydrodynamical simulations which included chemical evolution. These authors confirmed  the flattening of  metallicity gradients due to the infall of metal-poor gas generated  as the galaxies  approach each other. However, it was also reported that the triggering of star formation  by the increase of the central gas density could lead to the negative steepness of the abundance profiles as a consequence  of the new injection of  $\alpha$-elements  by  Type II supernovae (SNeII). After that, the overall gradient might get  flatter again due to a combination of diverse processes such as the ejection of metal-enriched gas from the central regions, the accretion of metal-rich  gas  in the outer parts via galactic fountains or tidal stripping  from  the companion galaxy \citep[see figure 7 in ][]{perez2011}.

Within a cosmological context, 
\citet{pilkington2012} reported the analysis of  the metallicity distribution in four  disc  galaxies with stellar masses in the range $2.7-5.9 \times 10^{10}$M$_{\odot}$, using different codes with 
a variety of subgrid physics and two analytical models. These authors found that the simulated galaxies exhibited negative metallicity gradients which were produced by the inside-out formation
of the disc, modulated by the characteristics of the subgrid physics \citep[see also][]{calura2012}. 
More recently, the evolution of the metallicity gradients with redshift was also  claimed as an observational feature which could be used to set constrains on the strength of the SN  feedback \citep{gibson2013,snaith2013}.

In this paper, we analyse the gas-phase oxygen  abundance profiles of the disc components of galaxies selected from a cosmological hydrodynamical simulation which includes the effects of a physically-motivated SN feedback and chemical evolution by SNeII and type Ia SNe (SNeIa).  The analysis is  carried out in the redshift range $z \sim [0,2.5]$.
Our aim is to study the abundance profiles
of the gas component in the simulated discs and to analyse the possible existence of a relationship between the slopes of the metallicity profiles and the  sSFR \citep{stott2014}. 

This paper is organized as follows. In Section 2 we describe the main characteristics of the simulations and the galaxy sample. In Section 3 we discuss the metallicity gradients and
SFR  for simulated disc galaxies and confront them with observations. Our main findings are summarized in the conclusions.

\section{numerical experiments and simulated galaxies}

We analyse a cosmological simulation of a typical field region of the Universe consistent with  in $\Lambda$-Cold Dark Matter ($\Lambda$CDM) scenario 
with $\Omega_{\Lambda}=0.7$, $\Omega_{\rm m}=0.3$, $\Omega_{b}=0.04$, a normalization of the power spectrum of $\sigma_{8}=0.9$ and $H_{0}= 100 h \ {\rm km} \ {\rm s}^{-1}\ {\rm Mpc}^{-1}$, with $h=0.7$. 
The simulation was performed by using the code {\small P-GADGET-3}, an update of {\small P-GADGET-2 } \citep{springel2001, springel2005}, optimized for massive parallel simulations of highly inhomogeneous systems\footnote{The version of  {\small P-GADGET-3} used in this paper does not  consider modifications to improve the performance of standard SPH on standard hydrodynamical tests.  However, \citet{schaye2015}  reported  that the impact of such modifications on the results of  cosmological simulations is small compared 
to those produced by variations in the subgrid physics. Nevertheless, it is an important aspect which we plan to consider in the near future. }. This version of {\small P-GADGET-3} includes treatments for metal-dependent radiative cooling, stochastic star formation (SF), chemical enrichment, and the multiphase model for the ISM and the SN feedback scheme of \citet{scan05,scan06}. 
This SN feedback model is able to successfully trigger galactic mass-loaded winds without introducing mass-scale parameters or imparting kicks to particles,
 for example.  As a consequence,  galactic winds naturally adapt to the
potential well of the galaxies where star formation takes place.

The multiphase model for the ISM allows the coexistence, interpenetration and material exchange  between  the  hot, diffuse phase  and the cold, dense gas phase \citep{scan06,scan08}. 
Stars form in dense and cold gas clouds  and part of them ends their lives as SNe, injecting energy and chemical elements into the ISM. The thermodynamical and chemical changes are introduced on particle-by-particle basis and considering the
physical characteristics of its surrounding medium. The detail description and tests of the multiphase medium and the SN feedback are presented
in \citet{scan06}. Briefly, it is assumed that each SN event releases $7\times 10^{50}$ erg, which are distributed among the cold, dense and hot, diffuse phases.  We assume  that $50$ percent of the energy is  injected into the cold phase surrounding the stellar progenitor (the rest is injected into the surrounding hot phase). \citet{scan08} reported this value to provide the 
best description of the energy exchange with the ISM for their scheme. 
 
We use the chemical evolution model developed by \citet{mosconi2001}. This model considers the enrichment by SNeII and SNeIa adopting the yield prescriptions of \citet{WW95} and \citet{iwamoto1999}, respectively. 
 We inject  $\epsilon_c = 80 \%$ per cent of the synthesized chemical elements  into the cold phase (the remaning $20\%$ is distributed directly within the hot phase). Galactic winds are responsible for transporting metals out of the galaxies from the cold ISM into
the hot circumgalactic medium.  After testing different $\epsilon_c$ for the injection of chemical elements, we
found that this value   provides a  good description of  
of the  metallicity gradients  of  the stellar populations in the disc components of the galaxies  (Tissera et al. in preparation) when compared with
the observational results from CALIFA survey \citep{sanchezb2014}. And as we will discuss later, the metallicity gradients of the gas-phase discs are also within the observed range.

 The lifetimes  for SNeIa are randomly selected within the range  $[0.1, 1]$ Gyr. This model, albeit simple, is able to reproduce well mean chemical trends   (for example, \citet{tissera2013} showed the mean $\alpha$-abundances as a function of [Fe/H] in the stellar haloes  of Milky-Way mass haloes). Moreover,   \citet{jimenez2014} and \citet{jimenez2015}  compared the median chemical abundances generated by  this SNIa lifetime model and the 
Single Degenerated scenario for the delay timelife distribution,  finding similar averaged trends (within the estimated dispersion; see their figure 4). Recall that in chemo-hydrodynamical simulations both the energy and metal injections are done on particle-by-particle
basis. A given gas  particle does not know which type of galaxy it inhabits and hence there is no a priori knowledge of the star formation history as 
it is the case in analytical models.

 The simulated volume represents a cubic region with  $14~$Mpc a side,  resolved with $2 \times 230^3$ initial particles. The mass resolution is  $5.9\times 10^{6}$M$_{\odot}$ and $9.1\times 10^{5}$M$_{\odot}$ for the dark matter and initial gas particles, respectively, with a maximum gravitational softening of $0.7~{\rm kpc}$. The initial condition has been chosen to correspond to a typical region of the Universe with no massive group present (the largest haloes have  virial masses smaller than $\sim 10^{13}M_{\odot}$). 
 \citet{derossi2013} compared the mass growth of haloes of different masses in a simulation performed with the same initial conditions  with those estimated by \citet{fakhouri2010} where a Halo Matching  Technique was applied to the Millennium Simulation. From this comparison, \citet{derossi2013}  showed that, at least, for this initial condition the growth of the haloes  is globally well-described
down to virial haloes $\sim 10^{10}$M$_{\odot}$. 
 Finally, \citet{pedrosa2015b} analysed the angular momentum content of the
discs and bulges of galaxies in this simulation, finding that the combination of star formation and feedback parameters
produced systems  which can reproduce  observational trends \citep[see][for results with a different subgrid parameteres]{pedrosa2014}.

\subsection{The simulated Galaxies}

We identified virialized structures by using a Friends-of-Friend algorithm and  then the  substructures were identified using the SUBFIND program \citep{springel2001}, which iteratively determines the self-bound substructures within the virial haloes. The specific angular momentum content and the binding energy are used to separate  baryonic particles dominated by rotation (i.e. disc component) from those dominated by dispersion (i.e. bulge components) as described in details by  \citet[][]{tissera2012}. 
 Our  simulated galaxies have angular momentum content and scale-lengths in agreement with observations \citep{pedrosa2015a, pedrosa2015b}.
In this simulation, the disc components of a  Milky-Way mass-sized galaxies  are resolved with $\sim 80000$ particles. A lower limit of  3000 baryonic particle in the disc components has been imposed to prevent strong resolution problems. Because of this  criterion, the number of galaxies analysed decreases for higher redshifts (see Table~\ref{table1}).

 To compare with the observational data, a characteristic radius ($r_{\rm gal}$) is defined to  enclose $\sim 83$ per cent of the stellar mass of the galaxy \citep[e.g.][]{tissera2012}. The metallicity  profiles are defined  by using the chemical abundances stored in the stellar or gas particles in the disc 
components. We adopted a maximum radius of $1.2r_{\rm gal}$ to avoid the outskirts of the discs which could be noisier due to the low particle density and/or   more affected
by nearby satellites. The inner cut-off radius is taken at $1.5~$kpc which is approximately $1.5$ times the maximum physical gravitational softening.

The simulated galaxies are  analysed in three redshift intervals around $z\sim 0, 1$ and  $2$, using the available snapshots.
 In each redshift interval,  galaxies are separated into two subsamples according to their stellar masses.  We adopt a stellar-mass limit  of M$_{ \rm star} = 10^{10}$M$\odot$. The properties of the simulated galaxies are summarized in Table~\ref{table1} and Table~\ref{table2}.

\begin{figure*}
\resizebox{5.3cm}{!}{\includegraphics{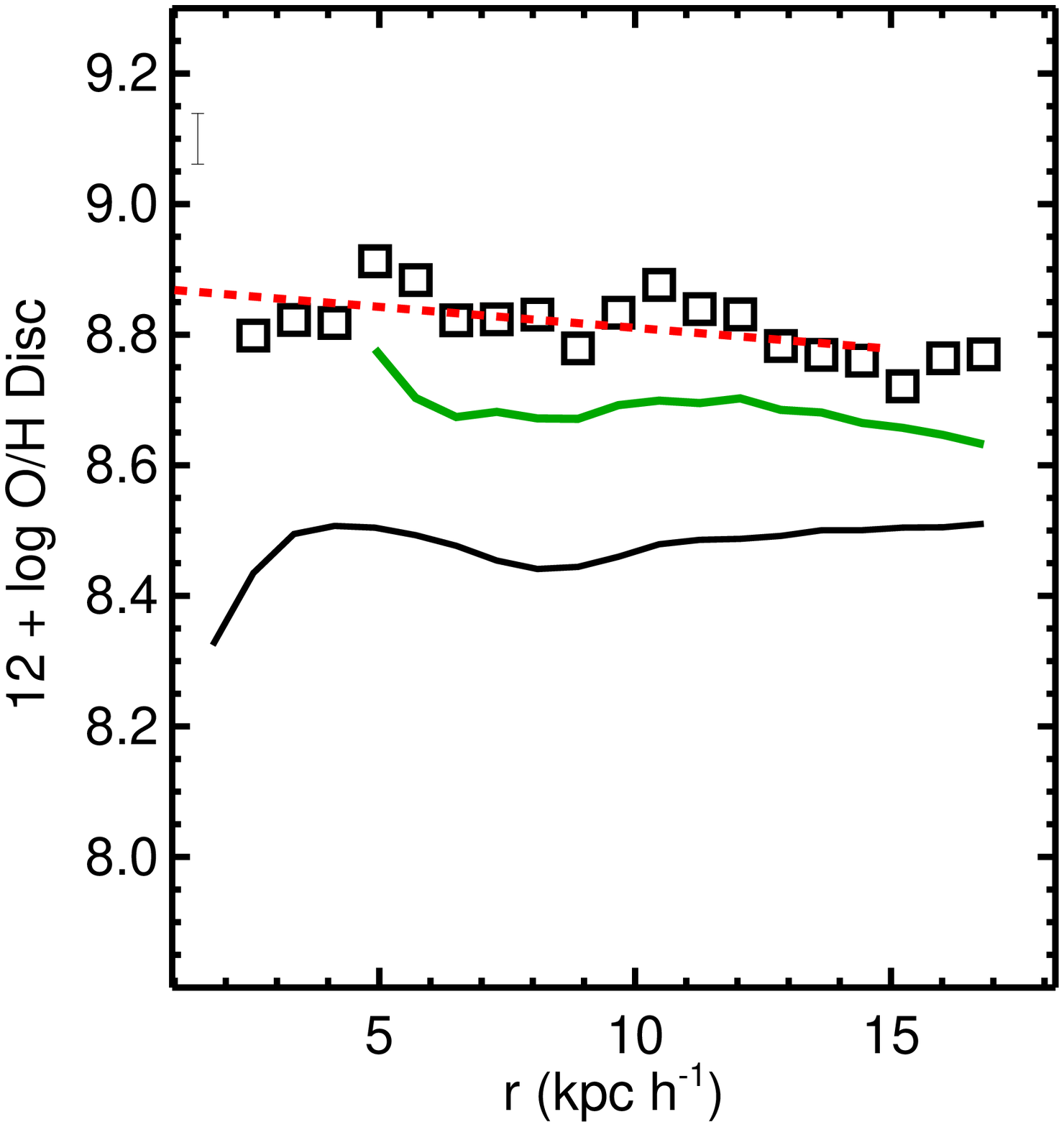}}
\resizebox{5.3cm}{!}{\includegraphics{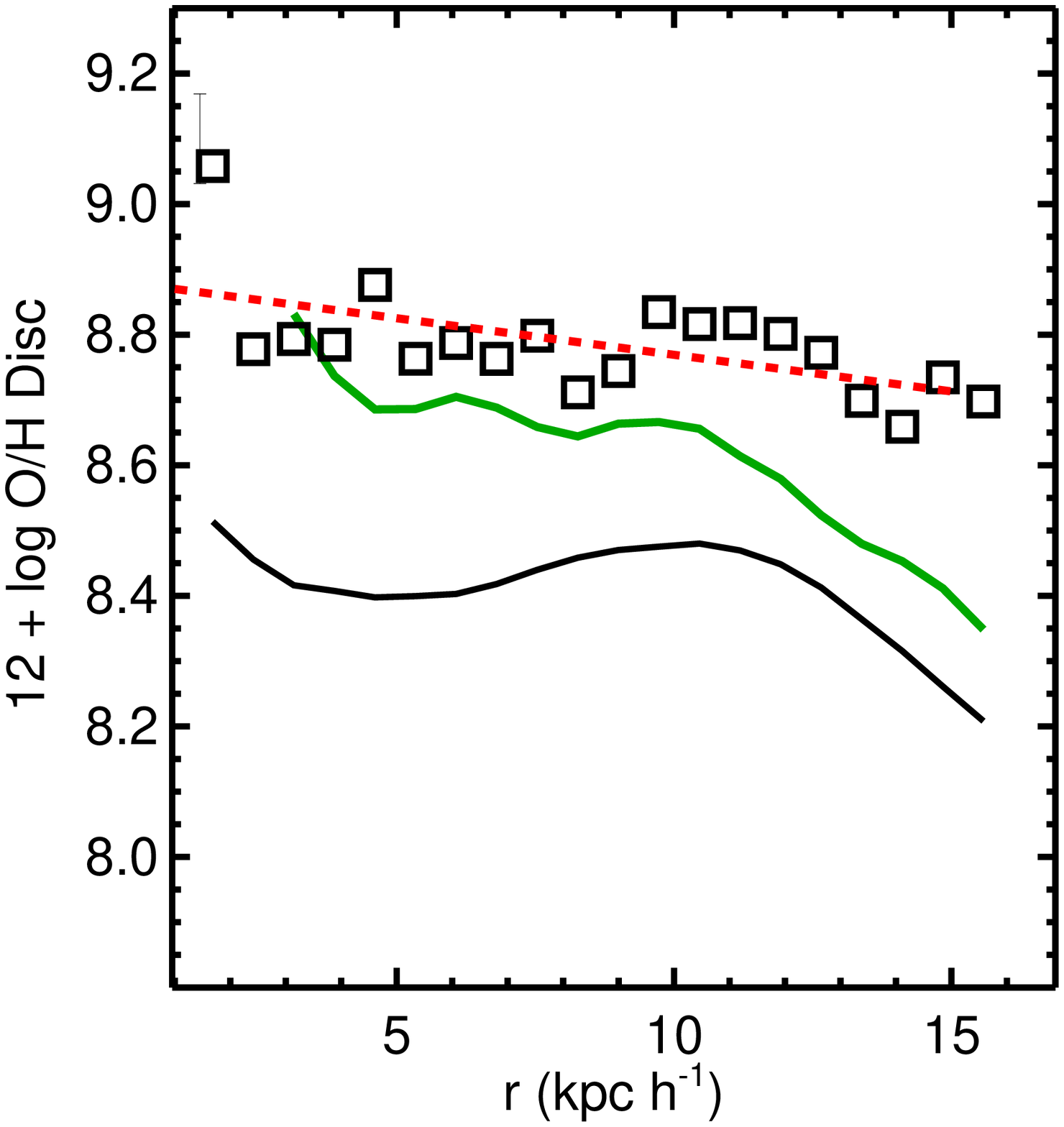}}
\resizebox{5.3cm}{!}{\includegraphics{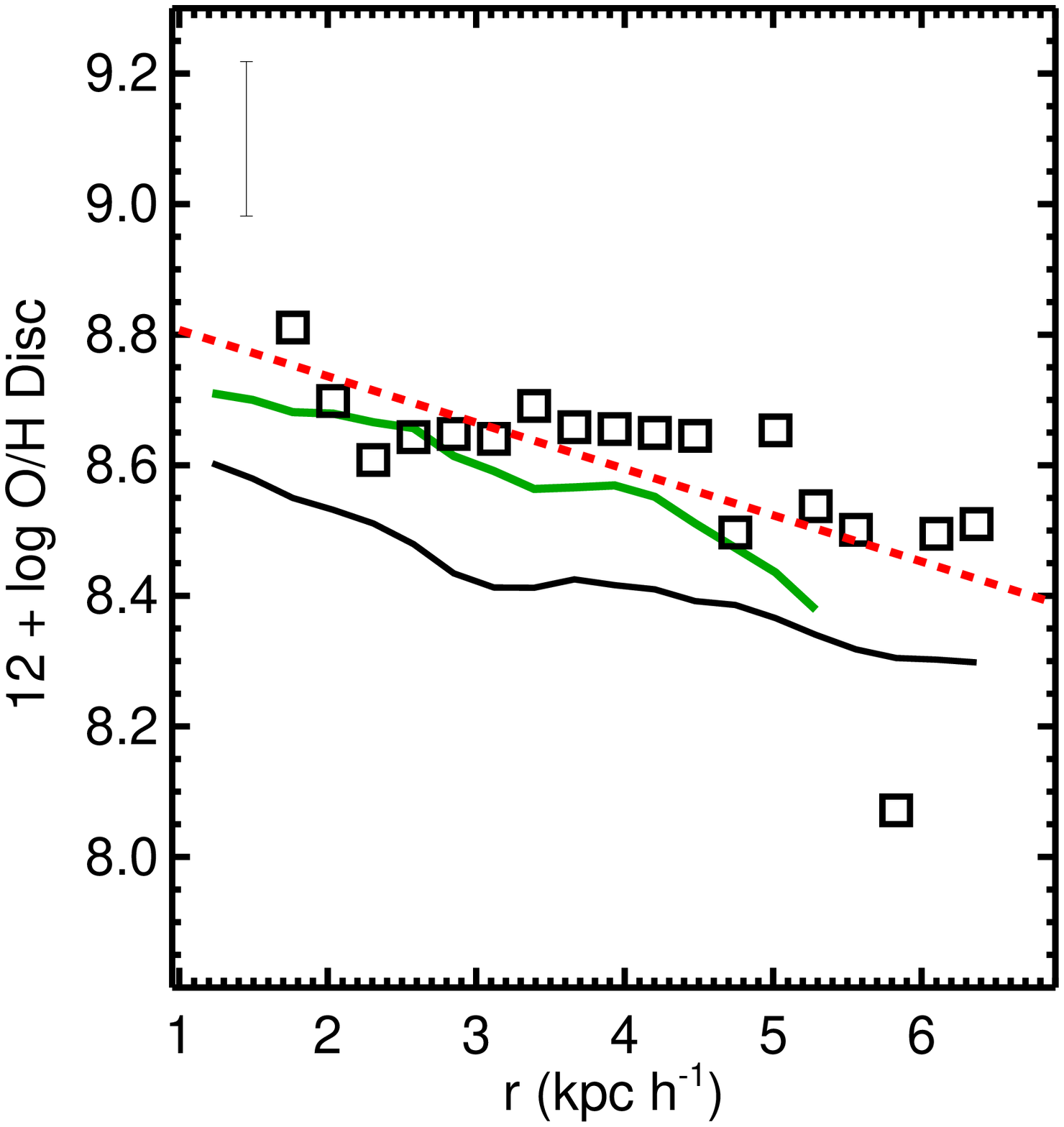}}\\
\hspace*{-0.2cm}
\caption{ Oxygen abundance profiles determined for  the gas component (open squares), the  total stellar disc population (solid black lines) and the stars younger than 2~Gyrs  in the simulated discs (green lines).
The linear regressions to the gas abundance profiles are also included (red dashed lines). The abundance gradients correspond to three  disc systems identified at $z\sim 0$ which exhibit negative abundance profiles. The total stellar mass of the galaxies (within $r_{\rm gal}$) are $6.68, 4.19$ and  $1.82  \times 10^{10}$ M$_\odot$ (see the Section 3 for more details). The error bars represent the mean rms of the linear fits. This set of systems has been chosen  only for illustration purposes.}
\label{gradients}
\end{figure*}

\section{Metallicity gradients and star formation of the disc galaxies}


The abundance profiles for both the gas and the stars in the disc components are estimated by using the  12+logO/H  ratios.
 We  calculated the median abundance values for the gas phase,  the total stellar populations and  those stars with ages smaller than 2~Gyrs, an age limit usually assumed to select young stellar population by observers. 

 The median
 abundance profiles for the gas and stellar components  are fitted with linear regressions (by a minimizing absolute deviation). 
 For illustration purposes, in  Fig.~\ref{gradients} we show the estimated gradients for three disc galaxies at $z \sim 0$,  with total  stellar masses (disc plus bulge stellar masses) between $\sim~2\times 10^{10}$ and $\sim~7\times 10^{10}$~M$_{\odot}$. 
As can be appreciated from this figure,  the gas components show close to flat or negative abundance gradients and are enriched with respect to the median abundances of 
the whole stellar populations on the discs, as expected. For these galaxies the oxygen slopes are $-0.004$, $-0.008$ and $-0.05~{\rm dex~kpc^{-1}}$ (from the  left panel to right one).
These values of slopes are consistent with the best abundance determinations available, based on direct measurements of the gas electron temperature to derive O/H \citep[e.g.][and references therein.]{Berg2015}

 As we can see, the gas abundances are also slightly higher than those of the stars younger than   2~Gyrs. In most cases the trends defined by the young populations and the gas components are similar as expected, but the gas is also enriched by  stars than   2~Gyrs. Hence, assuming this age interval to select stars to trace the current metallicity level of the ISM might  lead to underestimations (and in some cases they might also show metallicity profiles with different  slopes).

A similar analysis has been performed
for  simulated galaxies in the three redshift intervals. As a result, we have the gas-phase oxygen profiles of the  disc components for galaxies with a variety of stellars masses at
different stages of evolution.
 Hereafter we investigate the relationship between the oxygen gradients and the star formation activity as a function of their stellar mass by using the two subsamples defined in Section 2.1.
The same stellar-mass limit was applied to the observational data. In the case of \citet{ho2015} we use the mean values reported in 
their paper (table 1).

\begin{figure}
\resizebox{8cm}{!}{\includegraphics{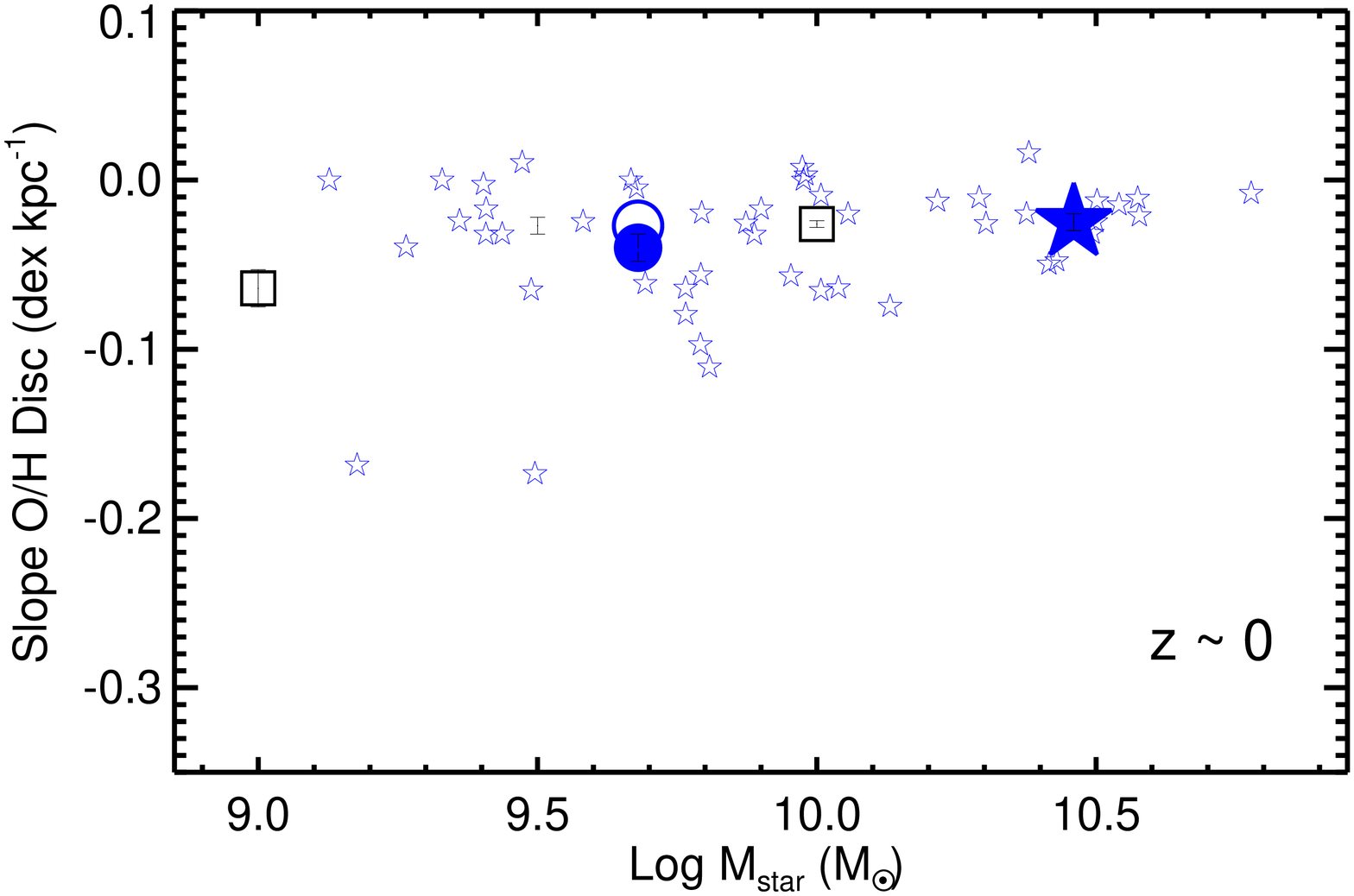}}
\resizebox{8cm}{!}{\includegraphics{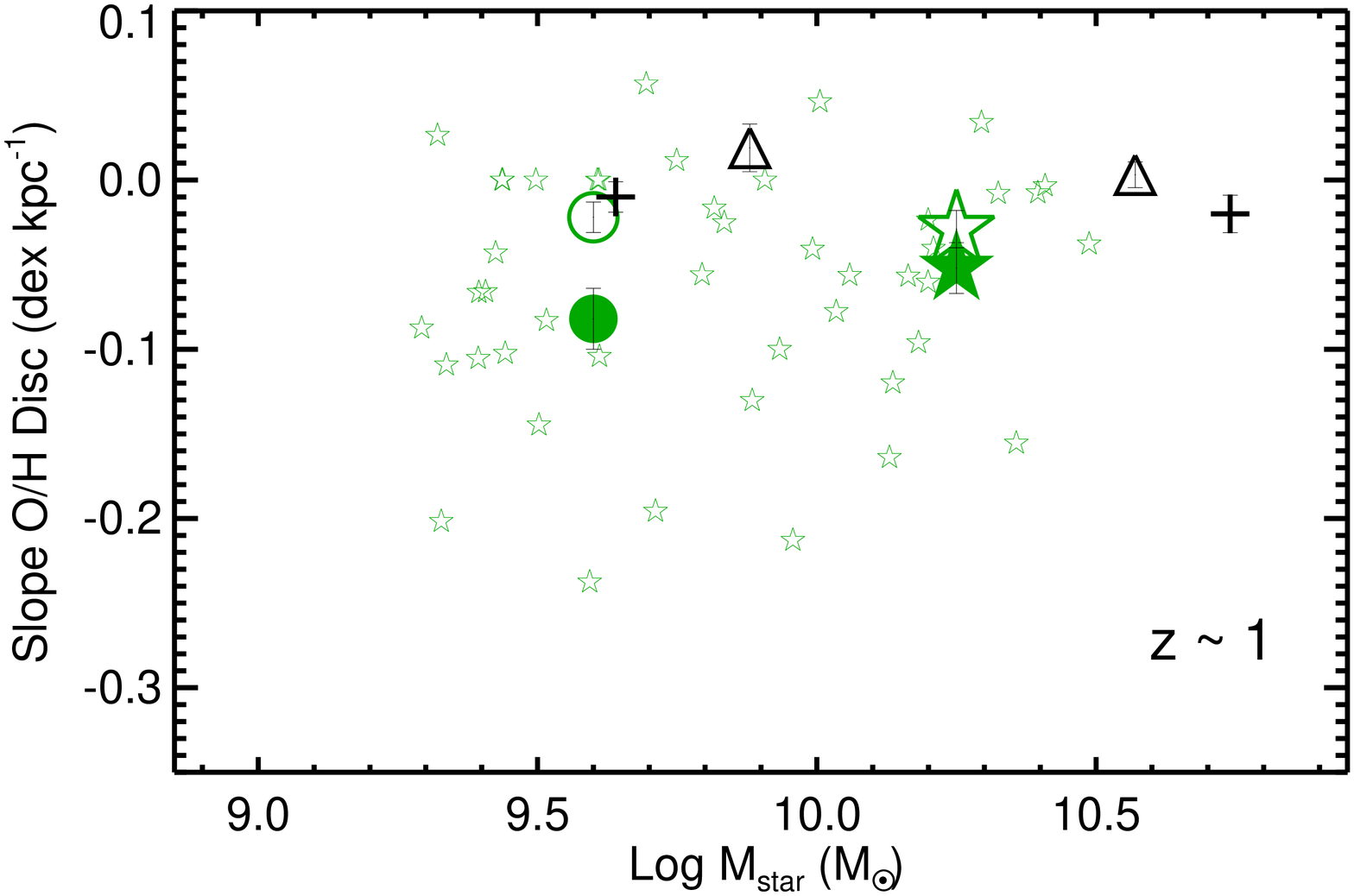}}
\resizebox{8cm}{!}{\includegraphics{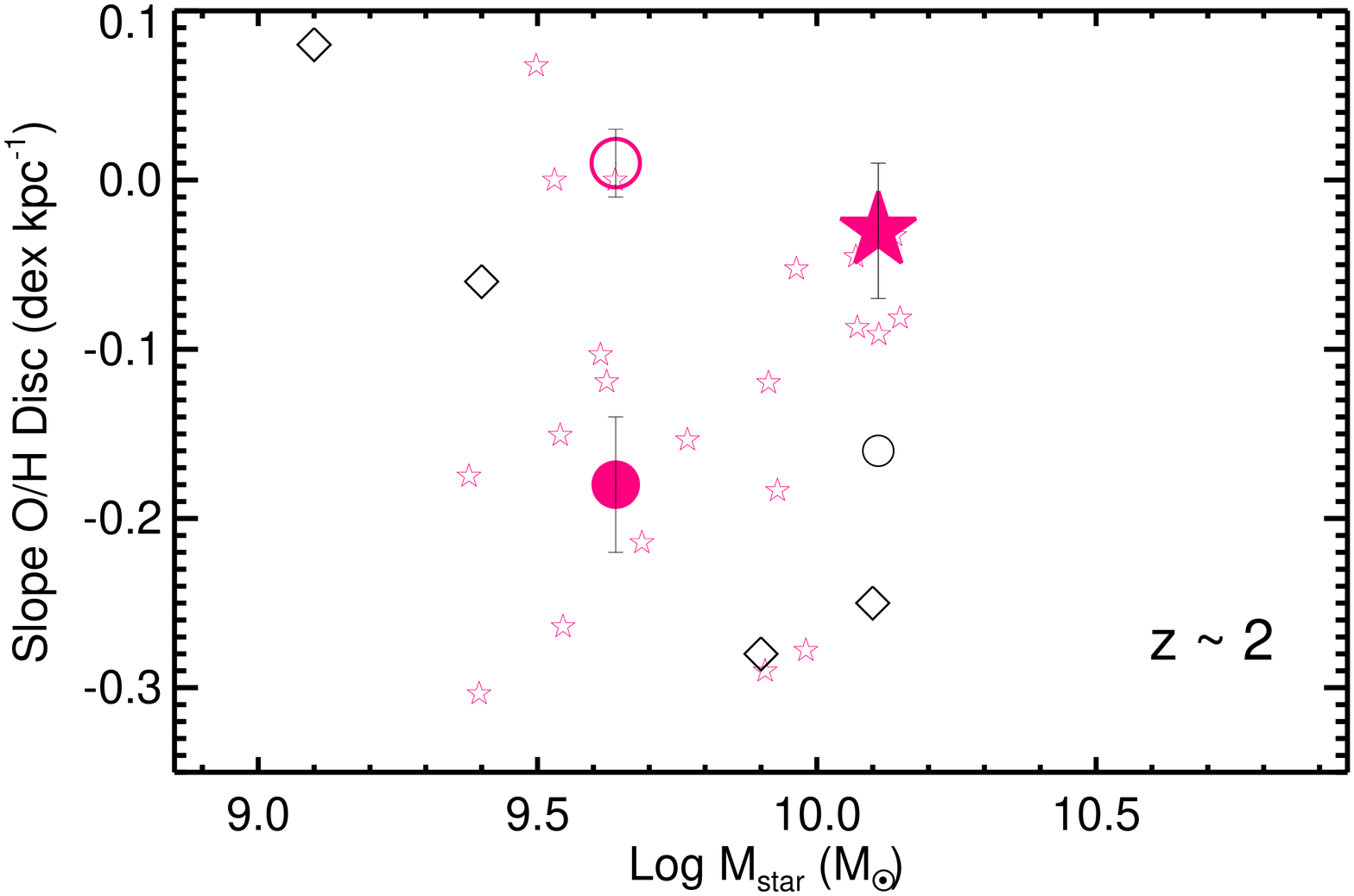}}
\caption{  Mean slopes of  gas-phase oxygen profiles  as a function of total stellar mass of the simulated galaxies in the low (large filled circles) and high (large filled stars) stellar-mass subsamples,  for $z\sim 0$ (blue symbols; upper panel), $z\sim 1$ (green symbols; middle panel) and $z\sim 2$ (magenta symbols; lower panel). 
Small open stars represent the values for individual simulated galaxies. The corresponding mean values obtained by excluding discs with gradients smaller than $-0.1~{\rm dex~kpc^{-1}}$  are also included for comparison (open symbols). 
 Mean observed slopes for disc galaxies are also included: \citet[black squares]{ho2015}  at $z\sim 0$ and  \citet[black triangles]{queyrel2012} and \citet[black  crosses]{stott2014} at $z\sim 1$ (see Section 3 for a detail description).  At higher redshift, individual metallicity slopes reported by  \citet[black  diamonds]{ jones2013} and  \citet[black  open circle]{yuan2011} have been included.
Mean values and error bars were estimated by applying a bootstrap technique.  
}
\label{masa_slope}
\end{figure}

In Fig.~\ref{masa_slope} we plot the  mean $M_{\rm star} $ and the mean oxygen slopes for the simulated  gaseous disc and the  observational data  as a function of redshift. For comparison, the values of the corresponding individual simulated galaxies are also included. The media values and the errors are estimated by using a bootstrap technique (Table~\ref{table1}). These errors are the {\it rms} obtained with respect to the mean bootstrap values.
As shown in  this figure,  at $z\sim 0$,  the simulated oxygen slopes show a trend with the stellar mass  which, albeit weak, is in agreement with the observational results reported by \citet{ho2015}.  Note that
the observational sample analysed by \citet{ho2015}  extends to lower stellar mass galaxies. 
The increase of the dispersion   in the metallicity gradients for galaxies with smaller stellar masses is also in agreement with  these observational results.
 At $z\sim 1$, the mean slopes are slightly more negative but  a similar trend with the stellar  mass is still present in the simulation.  However, observations from \citet{stott2014} and \citet{queyrel2012} exhibit more positive and flat metallicity gradients, with no clear correlation signal. Note that
the dispersion in the observations and simulated slopes are quite high  (the Spearman  rank correlation coefficients are $0.31 (0.03)$ and $0.32 (0.04)$ for  $z \sim 0$ and  $z \sim 1$, respectively.)

 At higher redshifts, there are few measures of  metallicity gradients and the available estimations show  large dispersion \citep{yuan2011,jones2013}.  Because of this, for the  $z \sim 2 $ interval, we plot the individual observed oxygen slopes instead of mean values as presented for the lower redshift ones.   The Spearman  rank correlation coefficient is $0.43  (0.05)$. Although it is slightly larger than those measured at lower redshift, the statistical significance is lower. 
Metallicity gradients smaller than $-0.1~{\rm dex~kpc^{-1}}$   are mainly
detected for galaxies with  low  stellar masses  at $z > 1$.  In fact,  galaxies with $M_{\rm star} > 10^{10} M_{\odot}$ exhibit mean oxygen slopes which do not evolve significantly since $z \sim 2$, on average.
The evolution of  the slopes of the metallicity profiles is  determined by simulated galaxies with  $M_{\rm star} < 10^{10} M_{\odot}$.

In order to assess the impact of disc galaxies with slopes more negative than  $-0.1~{\rm dex~kpc^{-1}}$  on this correlation, we estimated the mean properties by excluding them.  In this case, the mean values of the oxygen slopes of discs in the high stellar-mass subsample are weakly affected since they have small or null fractions 
of discs with very negative slopes (only at  $z\sim 1, $, we detected a fraction of $19\%$ of the high stellar-mass discs to have such low metallicity gradients). Conversely, the results for galaxies in the  low stellar-mass subsample are strongly modified.  If disc galaxies with  slopes smaller than $-0.1~{\rm dex~kpc^{-1}}$  are not considered, 
the mean slopes show either no correlation or a negative one, as it  can be seen  in  Fig.~\ref{masa_slope} (open symbols). The fraction of discs with oxygen slopes lower than $-0.1~{\rm dex~kpc^{-1}}$ increases with increasing redshift for galaxies in this subsample ($21\%, 40\% $ and $78\%$ for 
$z\sim 0,1,$ and 2, respectively).

We found no   relation between the slopes of the oxygen profiles and the number of particles used to resolve the disc components (i.e. low number discs exhibit positive as well as negative slopes). And since we have used the same minimum limit in the number of particles at all analysed redshifts, numerical effects should affect similarly the three redshift subsamples. 
The metallicity gradients show no clear correlation with the  gas fraction in the disc components.

\begin{table*}
	\centering
	\caption{Mean global properties of the simulated disc galaxies  in the two defined
stellar-mass subsamples  as a function of redshift ($z$).
 The third column gives the number (N) of disc galaxies in the corresponding subsample. The three following columns provide: the mean gas-phase oxygen gradients (dex~kpc$^{-1}$), the mean star formation rates ($M_\odot yr^{-1}$) and then mean specific star formation rates ($10^{-10} yr^{-1}$) for simulated discs galaxies.  The last two columns show the parameters f$_{<-0.1}$  and  f$_{>0}$ which represent the fractions of discs with  metallicity slopes smaller than $-0.1~{\rm dex~kpc^{-1}}$  and with positive metallicity slopes,  respectively.} 
	\begin{tabular}{lcccccccr} 
		\hline
                Redshift   & Log $M_{\rm stars}$      &N&   <Slope>	                   &   <SFR>	&    <sSFR> &    f$_{<-0.1}$  & f$_{>0}$ \\
                \hline
                    $z\sim 0$   &$ < 10$ &29 &-0.041 $\pm$ 0.008     &0.29 $\pm$ 0.05      & 0.7 $\pm $ 0.2  & 0.11 & 0.10\\
                                       &$  >10$ &20 &-0.025 $\pm$ 0.005     &1.85 $\pm$ 0.41       & 0.6 $\pm $ 0.1  & 0.0  & 0.05\\
                     $z \sim 1$ & $< 10$ &30 &-0.080 $\pm$ 0.018     &1.35 $\pm$ 0.29       & 4.04 $\pm $ 0.05 & 0.40 & 0.10\\
                                       & $>10$  &16 &-0.052 $\pm$ 0.015     &6.80 $\pm$ 0.91        & 3.67 $\pm $ 0.04 & 0.19 & 0.13\\
                     $z \sim 2$ & $< 10$ &18 &-0.18 $\pm$ 0.04         &4.79 $\pm$ 0.63        & 10.90 $\pm $ 1.50 &  0.78 & 0.05\\
                                       & $>10$  &6   &-0.03 $\pm$ 0.04         &7.32 $\pm$ 0.87         & 5.90 $\pm $ 0.60  & 0.0 & 0.17\\
		\hline
	\end{tabular}
\label{table1}
\end{table*}

\begin{table}
	\centering
	\caption{Properties of the simulated disc galaxies. 
The stellar mass  $M_{\rm stars}$ (second column), the star formation rate ($M_\odot yr^{-1}$; third column), the  specific star formation rate ($ yr^{-1}$, fourth column) and the gas-phase oxygen gradient (dex~kpc$^{-1}$; fifth column) are given for ten  simulated discs galaxies.  The complete table is available online.} 
	\begin{tabular}{lcccc} 
		\hline
                Redshift   &                                        Log $M_{\rm stars}$    	                   &   SFR	&    sSFR  & Slope\\
                \hline
                                                                 & 10.98  &   7.91  &8.30e-11 & -0.005\\
                                                                 &  10.78   &   4.46 & 7.44e-11  &-0.008\\
                                                                 &10.56   & 2.76   &7.31e-11   &-0.021\\
                                                                 &10.42     & 2.55   & 9.70e-11  & -0.050\\
       \multirow{3}{*}{$ z \sim  0 $}             & 10.57     & 2.62  &6.99e-11   &-0.011\\
                                                                    &10.50     &0.99  & 3.15e-11  & -0.027\\
                                                                  & 10.43   & 0.84  & 3.14e-11  & -0.048\\
                                                                  & 10.29    & 0.79  &4.02e-11   &-0.010\\
                                                                  & 10.38    & 0.85  &3.57e-11   &-0.020\\
                                                                  & 10.30   & 1.09  &5.44e-11   &-0.026\\
     		\hline
	\end{tabular}
\label{table2}
\end{table}

\begin{figure}
\resizebox{8cm}{!}{\includegraphics{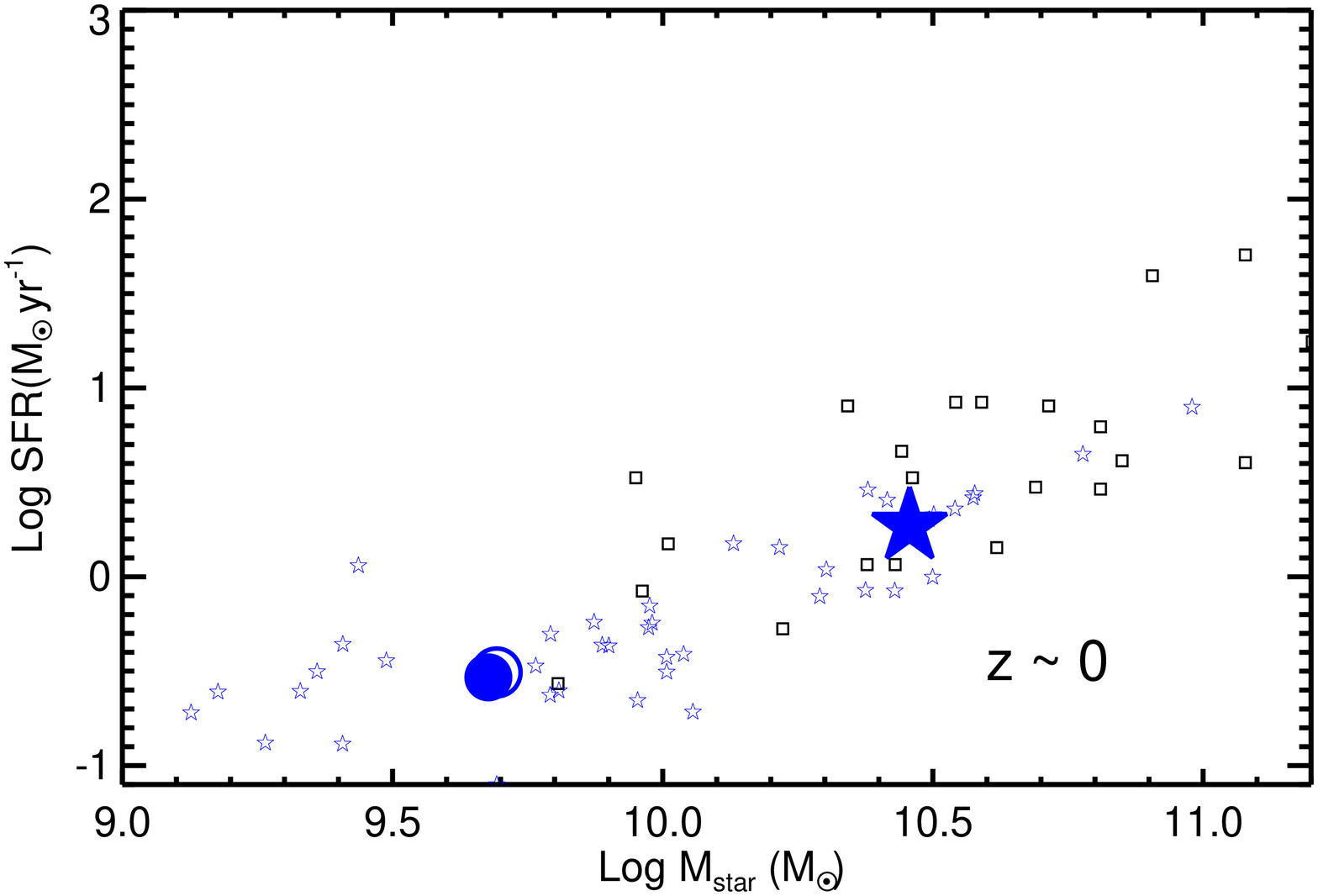}}
\resizebox{8cm}{!}{\includegraphics{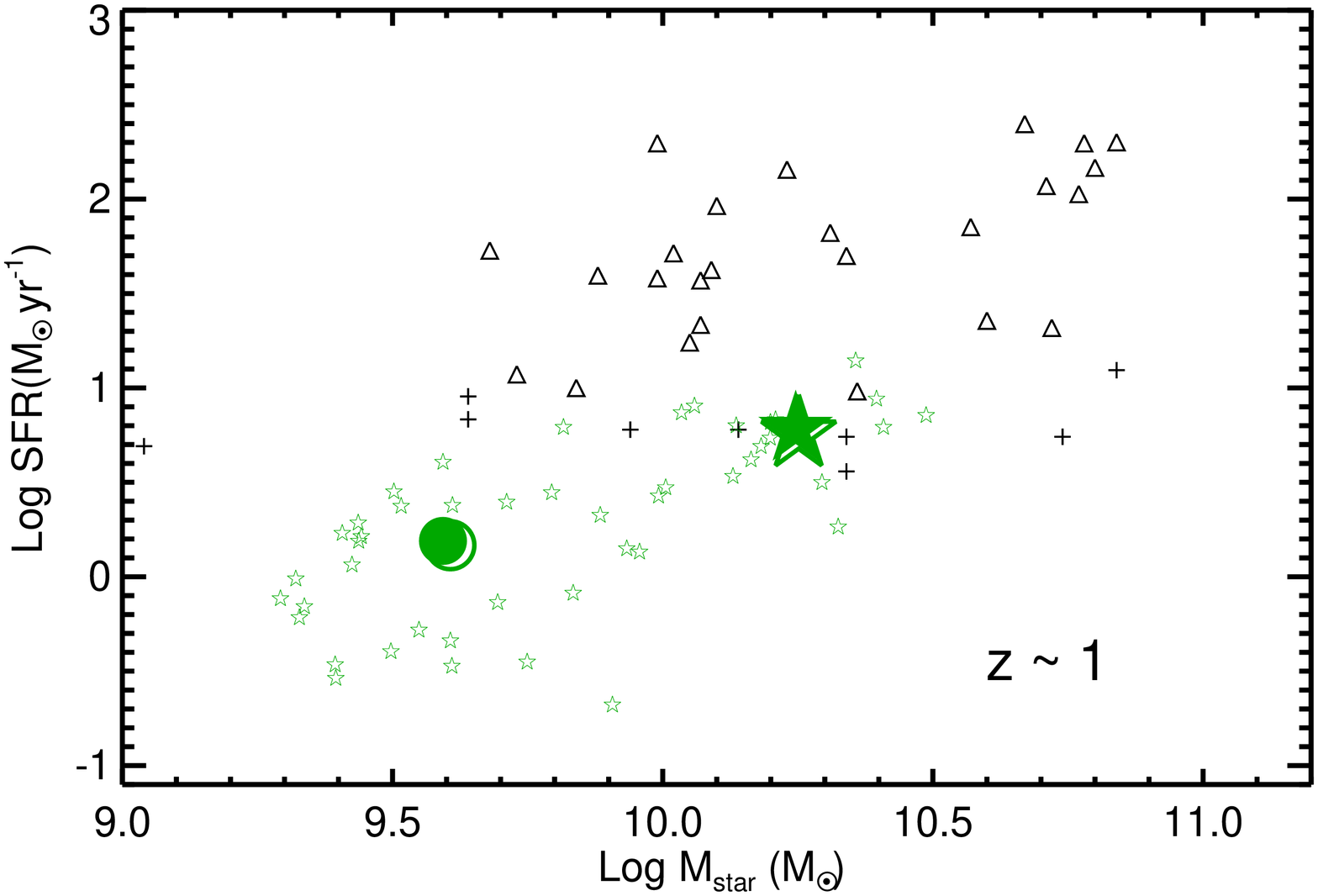}}
\resizebox{8cm}{!}{\includegraphics{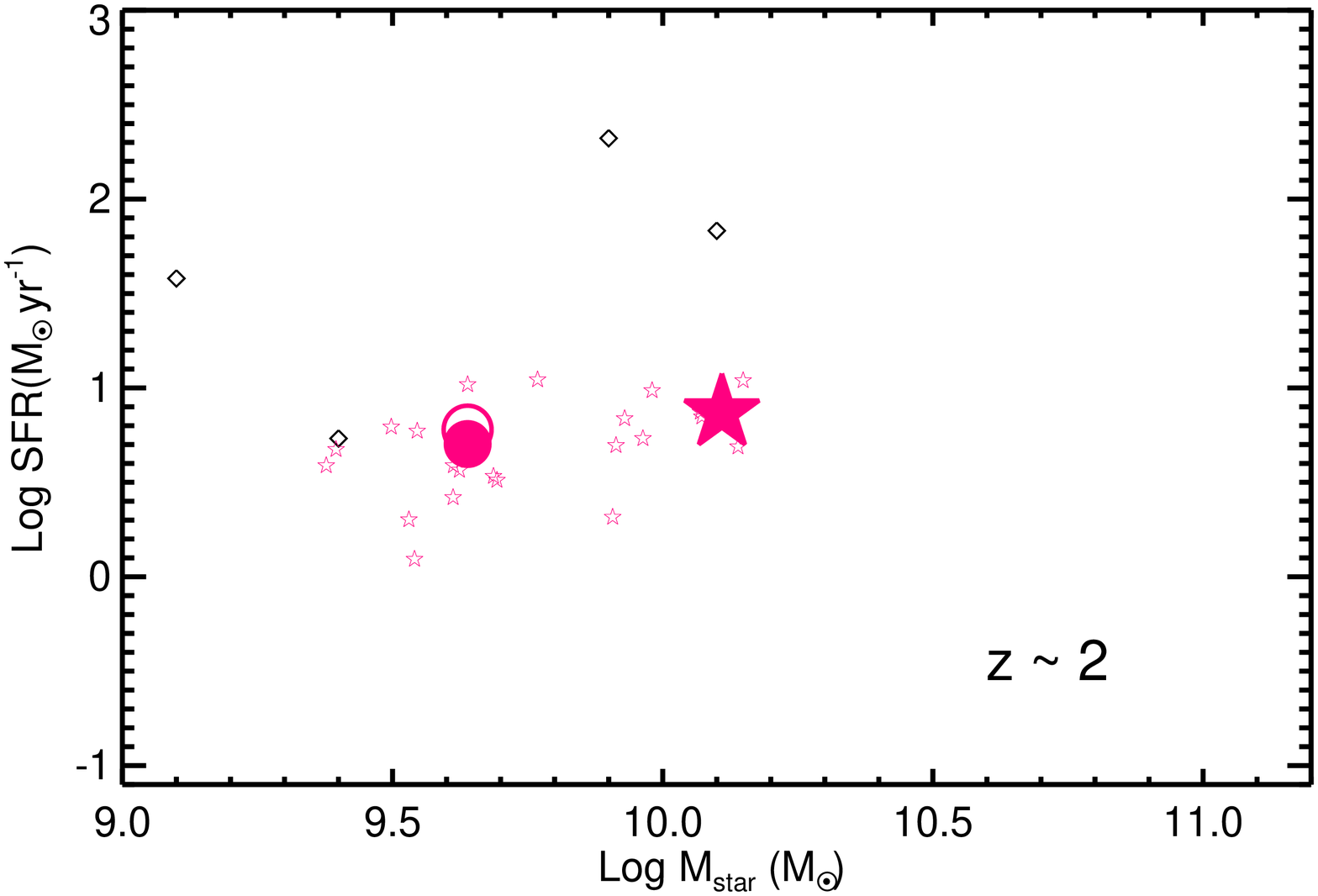}}
\hspace*{-0.2cm}
\caption{  Mean star formation rate as a function of the mean stellar mass of simulated galaxies  in the low (large filled circles) and high (large filled stars) stellar-mass subsamples for  $z\sim 0$  (blue symbols;  upper panel),  $z\sim 1$ (green symbols;  middle panel),  $z\sim 2$ (magenta symbols; lower panel). Small open stars represent the values for individual simulated galaxies. Mean estimations obtained by excluding galaxies with slopes $<-0.1~{\rm dex~kpc^{-1}}$ are also included (open symbols). Most of them are superposed to the mean values from the whole sample. For comparison, we  included the observational results from   \citet[][black squares; $z \sim 0$]{rupke2010}, \citet[black triangles; $z \sim 1$]{queyrel2012},  \citet[black, crosses; $z \sim 1$]{stott2014} and \citet[black, diamonds; $z \sim 2$]{jones2013}.
Mean values and error bars were estimated by applying a 
bootstrap technique.}
\label{masa_sfr}
\end{figure}

In order to understand the origin of both positive and very negative metallicity gradients, we first investigate if these galaxies
follow the observed correlation between SFR and stellar mass
 \citep{karim2011} as a function of redshift 
\citep[][]{daddi2007}.  In Fig. ~\ref{masa_sfr} we show   the mean SFRs  and the  mean stellar masses of  the simulated galaxies in the two stellar-mass subsamples
as a function of redshift. 
The simulated relations are consistent with the observed one  at $z \sim 0$. For higher redshifts, the simulated galaxies tend to have lower SFR than observations. However, we note that, on one hand, the range
of stellar masses covered by the simulated galaxies at $z >0$ is smaller than the observed ones due our small simulated volume (note that the mean values for the high stellar-mass subsamples move to lower masses at higher redshifts).  
And that, on the other hand,  observations tend to survey higher star formation systems for higher redshift
 \citep[e.g.][]{mannucci2010,salim2014} while in the simulations we are including galaxies of all types. 
Hence, the comparison should be taken with caution.
Overall we find that at a  given stellar mass, the star formation rate is larger for higher redshift.

\subsection{ A possible relation between the metallicity gradient and the sSFR? }

\begin{figure*}
\vspace{-0.6cm}
\resizebox{15cm}{!}{\includegraphics{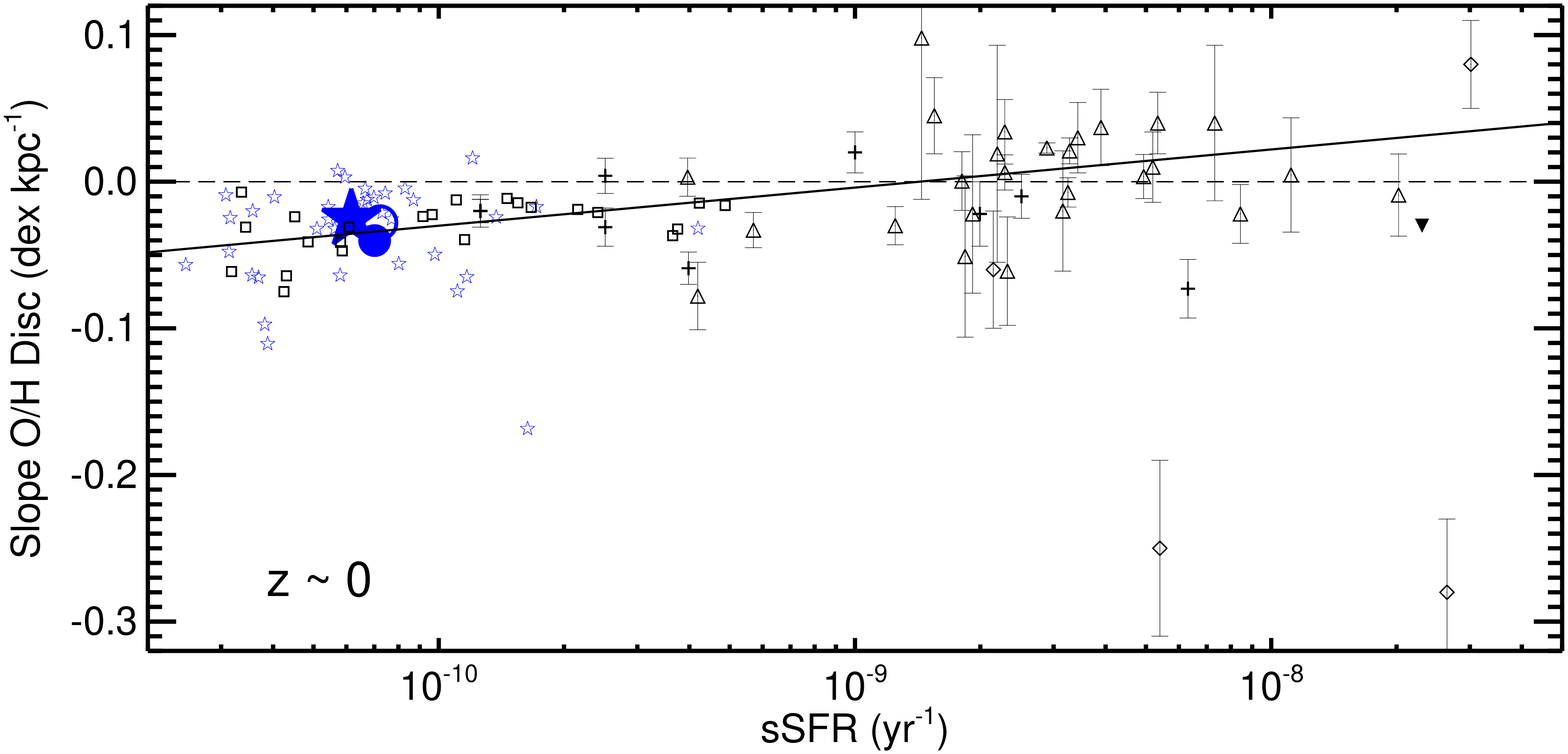}}
\vspace{-0.6cm}
\resizebox{15cm}{!}{\includegraphics{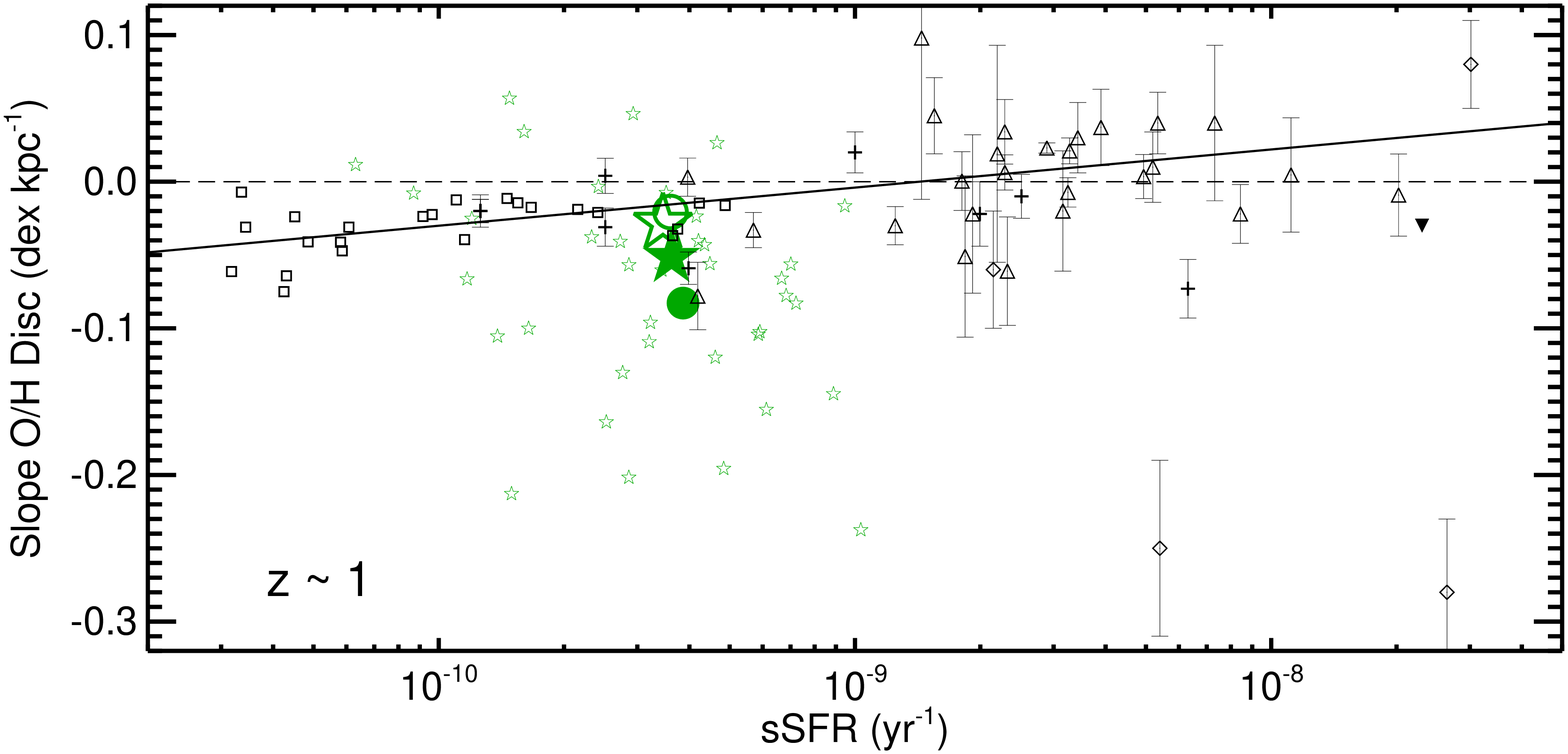}}
\vspace{-0.6cm}
\resizebox{15cm}{!}{\includegraphics{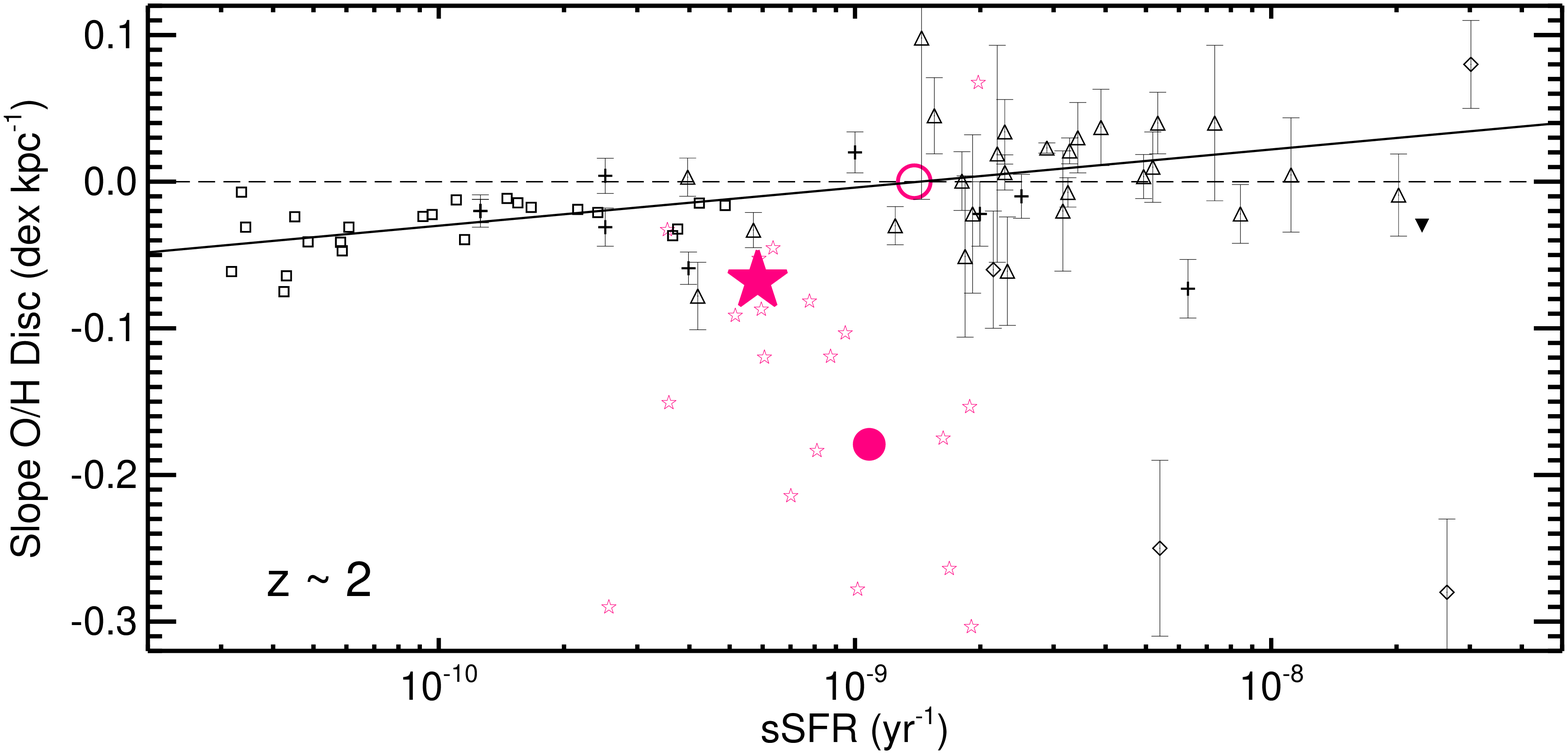}}
\hspace*{-0.2cm}
\caption{ {Mean slope of the  gas-phased oxygen abundance profiles for  the simulated ISMs as a function of the mean sSFR for galaxies  for $z \sim 0$ (upper panel), $z \sim 1$ (middle panel)
and $z \sim 2$ (lower panel) .
Mean values have been estimated for galaxies in the low and high stellar-mass subsamples (large filled circles and stars, respectively). Small open stars represent the values for individual simulated galaxies (see  Fig.~\ref{masa_slope} for color code)}. Estimations obtained by excluding oxygen slopes smaller than $-0.1~{\rm dex~kpc^{-1}}$ are also displayed (open circles and stars, respectively).  For comparison we include the observational results from  \citet[black, open squares]{rupke2010}, \citet[black, open triangles]{queyrel2012}, \citet[black, open rombus]{jones2013} \citet[black, inverted triangles]{jones2015}, \citet[black, crosses]{stott2014}  and the linear regression  reported by \citet[solid line]{stott2014}.  }
\label{slopessfr}
\end{figure*}

Recently \citet{stott2014} reported evidences of a possible correlation between the slope of the oxygen gradients of disc galaxies and their sSFRs.  These authors  estimated these
properties in their galaxy sample at $z\sim 1$ and included  observations available in the literature for other redshifts. 
 They proposed that mergers, and possibly secular evolution, could provide an explanation of the reported trends since  they are known to be efficient mechanisms to drive gas inflows.
If the galaxies have  negative metallicity gradients, then these gas inflows will contribute with low metal-enriched gas, diluting the central abundances.
 In order to assess if such correlation is also present in our simulated galaxies, in  Fig.~\ref{slopessfr}
 we plot the mean gas-phase metallicity slopes versus mean sSFR  for the simulated galaxies in  the two defined stellar-mass subsamples. Estimations for the the three defined redshift intervals
have been included.

As can be seen from  Fig.~\ref{slopessfr}, at $z\sim 0$ simulated galaxies with low and high stellar-masses are consistent with the observed relation reported by \citet{stott2014}.
In particular, massive galaxies
have mean metallicity slopes and sSFRs which are consistent with the  observed relation at all analysed redshift.
For galaxies in the low stellar-mass subsample,  the variation  of the slopes and sSFRs with redshift are larger. These galaxies increase  their sSFR for more than order of magnitude between $z\sim 0$ and $z\sim 2$ while their mean metallicity slopes decrease more strongly. As a consequence, they deviate form the observed relation. 
In the $z\sim 2$ interval, simulated galaxies show a wide spread in slopes going from positive to
very negative ones. This is also true for the available observations.

We found that independently of their stellar masses, simulated galaxies have larger diversity of metallicity gradients with increasing redshift (Table~\ref{table1}).  The fraction of galaxies with positive metallicity slopes increases with increasing redshift from $\sim 5$ percent at $z\sim 0$ to $\sim 17$ per cent at $z\sim 2$ for massive galaxies while for those with lower stellar masses, the
fraction of positive metallicity slopes remains approximately constant with redshift ($10\%$).
Conversely, the fraction of disc galaxies with metallicity slopes smaller than $-0.1~{\rm dex~kpc^{-1}}$ is larger in the low stellar-mass subsample and  increases with redshift.

To evaluate the impact of the presence of the metallicity gradients smaller than  $-0.1~{\rm dex~kpc^{-1}}$, we re-estimated the  mean metallicity slopes and mean SFRs for the simulated galaxies  excluding
them. In this case, the mean simulated  values for 
both the low and the high stellar-mass subsample are consistent with the  Stott's relation (note that the two observed points of \citet{jones2013} had been  included in Stott's estimations).

 Hence it would be important to confirm observationally the existence of such steep negative gradients at high redshift and their frequency in order to make a robust comparison with the simulations. These observations will provide important constrains for the sub-grid physics (principally the SN feedback model).

\subsection{What are these systems?}

In order to understand the origin of the very negative and the positive slopes, we explore  the morphologies of the simulated
galaxies and estimate the distances to the nearby galaxies.
No major mergers are identified for these systems since $z < 3$ but they experienced minor mergers and interactions \citep{pedrosa2014}.

\begin{figure*}
\resizebox{4.3cm}{!}{\includegraphics{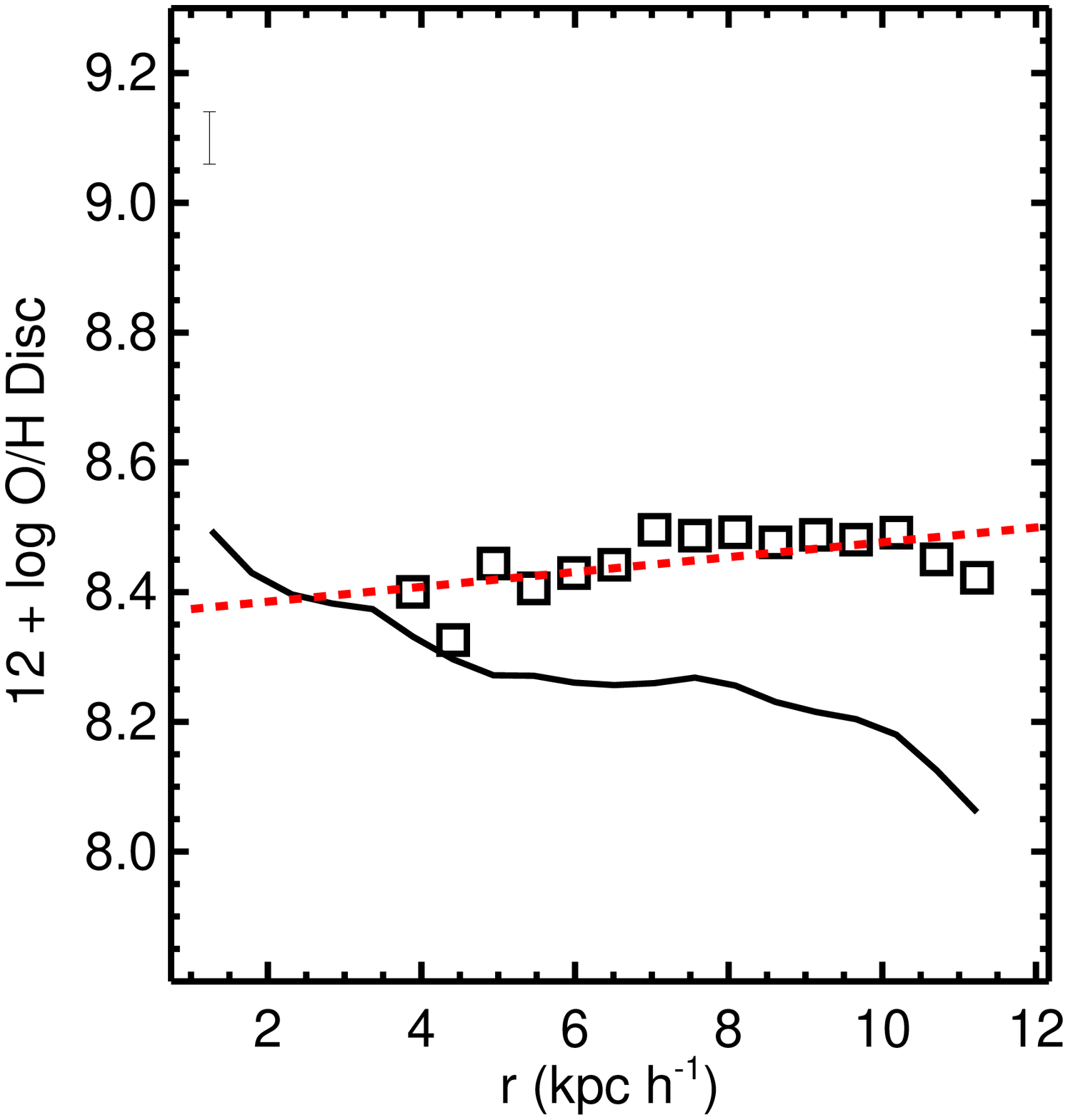}}
\resizebox{4.3cm}{!}{\includegraphics{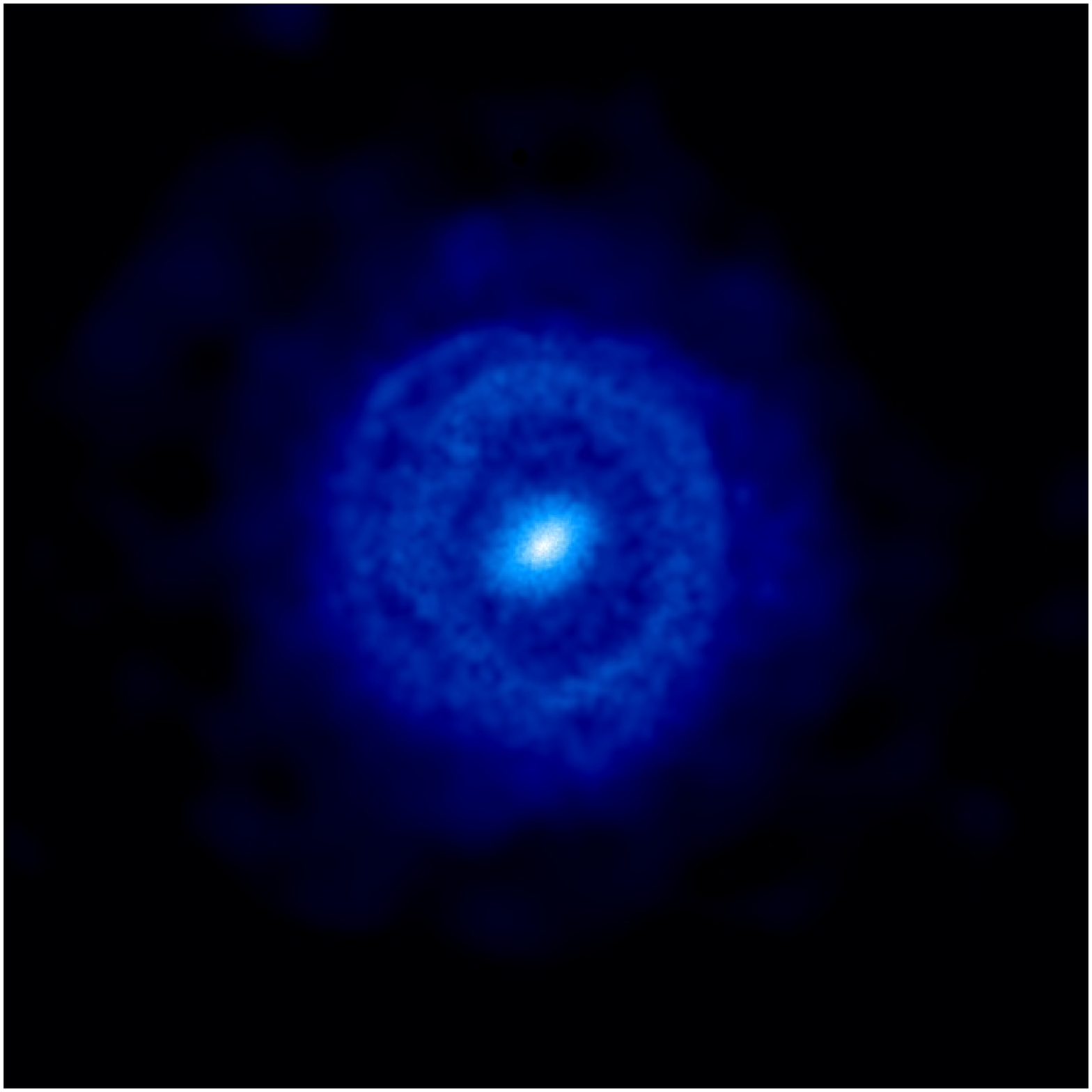}}
\resizebox{4.3cm}{!}{\includegraphics{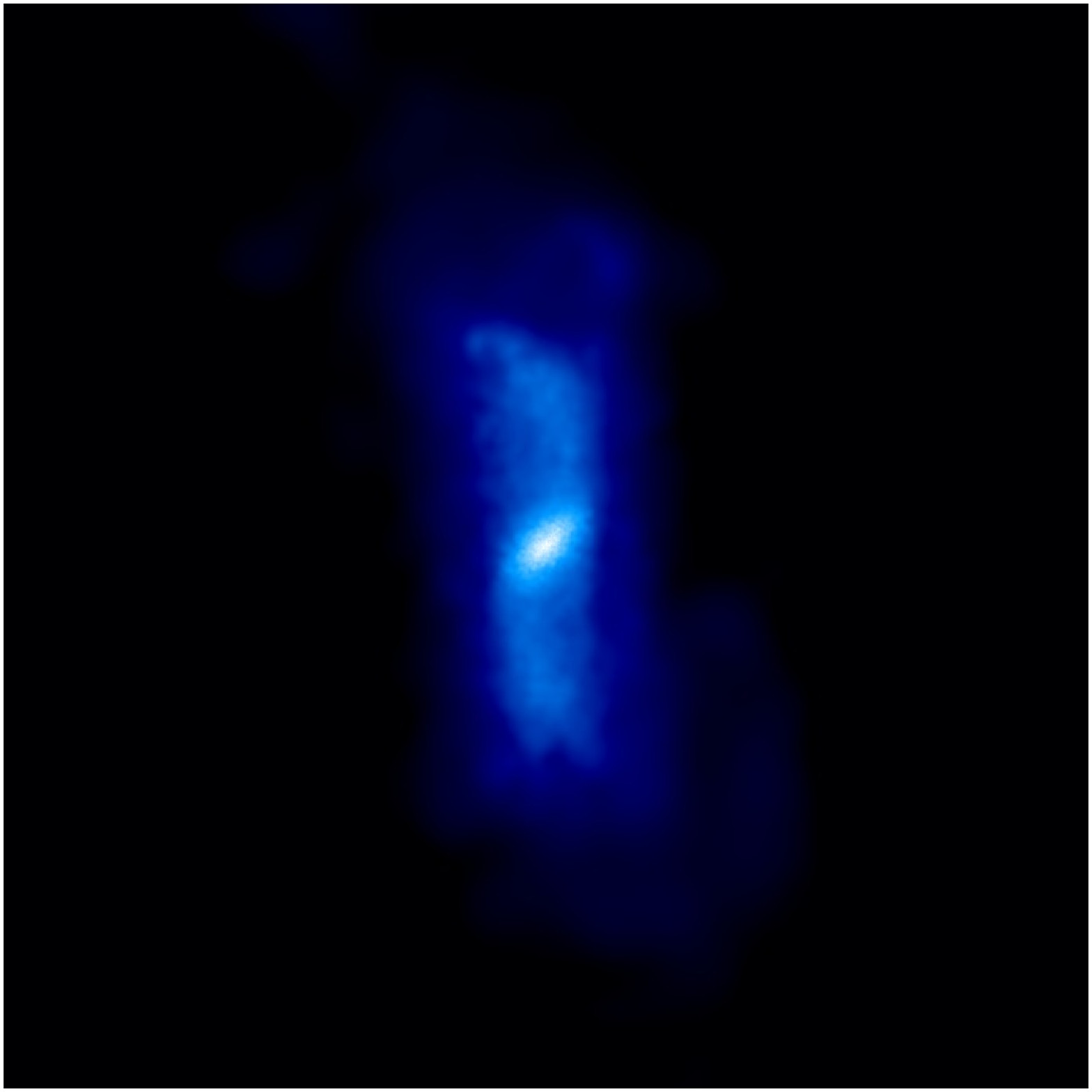}}
\resizebox{0.9cm}{!}{\includegraphics{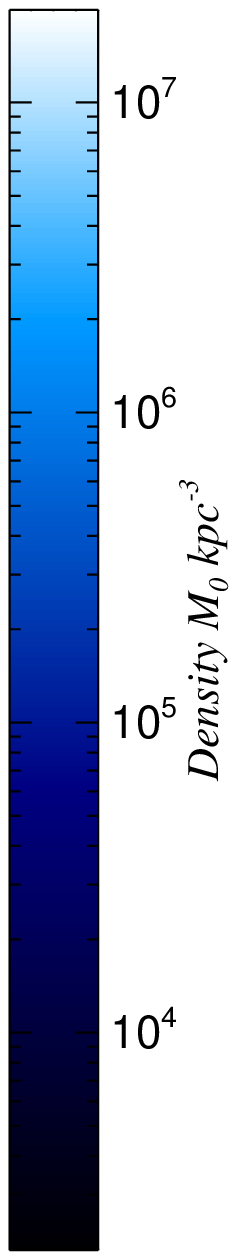}}\\
\resizebox{4.3cm}{!}{\includegraphics{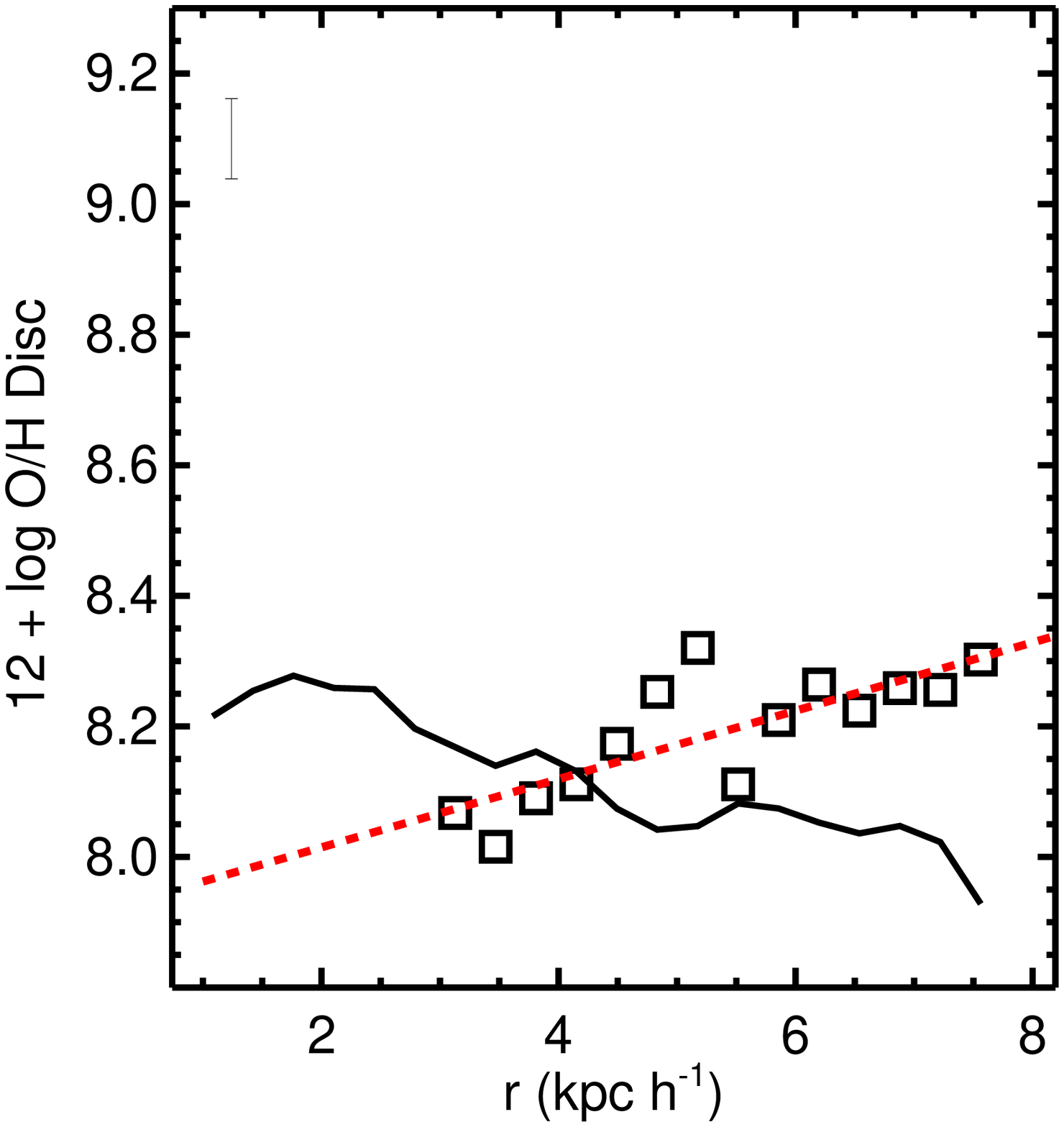}}
\resizebox{4.3cm}{!}{\includegraphics{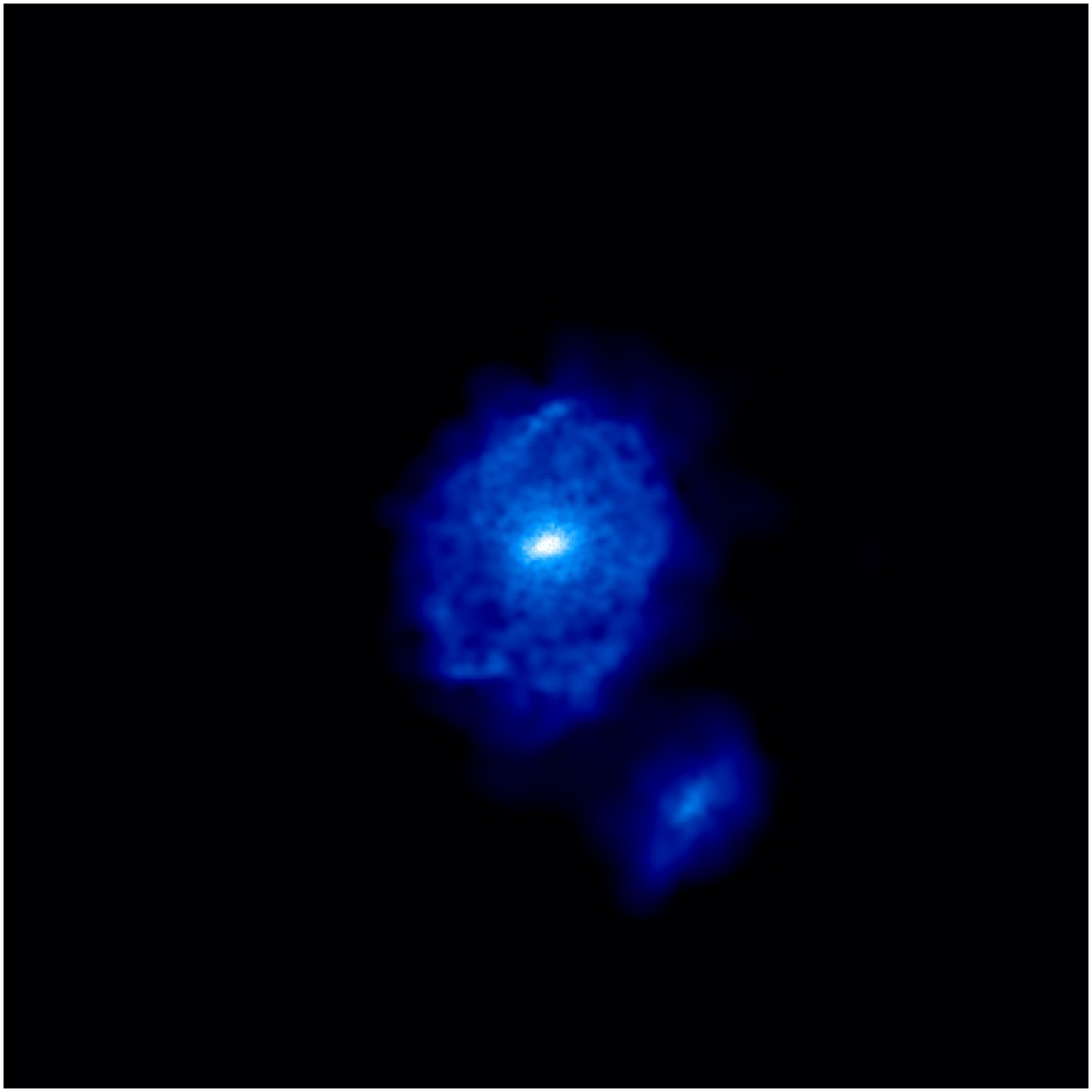}}
\resizebox{4.3cm}{!}{\includegraphics{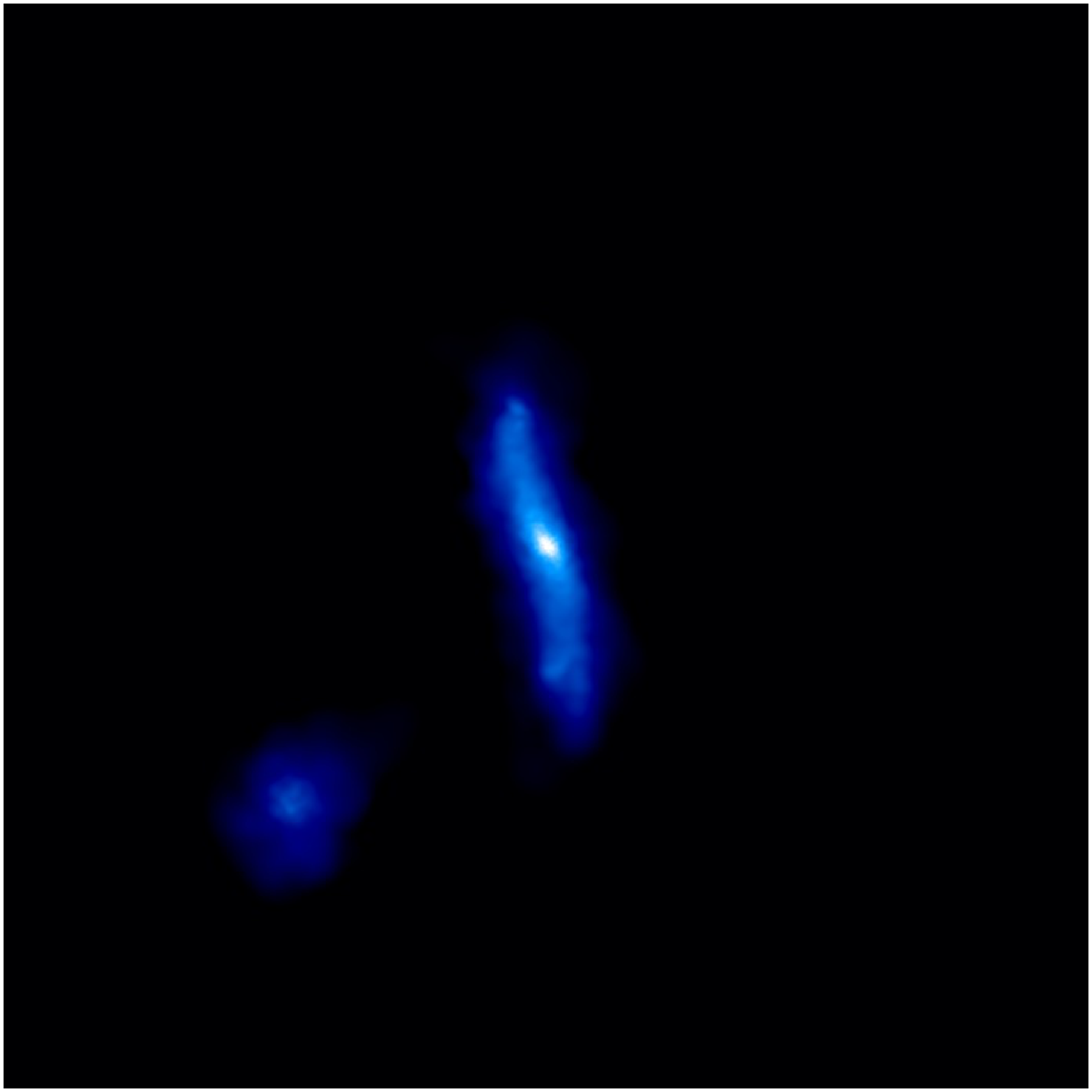}}
\resizebox{0.9cm}{!}{\includegraphics{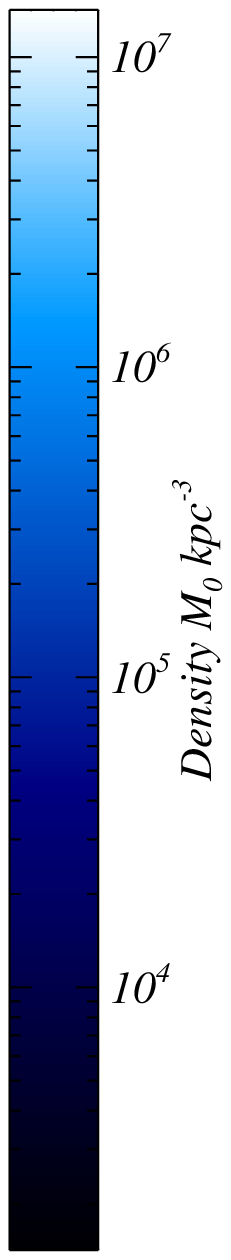}}
\caption{ Mean $12+$log~O/H profiles  determined for  the gas-phase disc components (open squares) and the  stellar discs populations (black solid curves) in  simulated  galaxies which show positive 
gas abundance profiles. 
The linear regressions to the gas abundance profiles are also included (red lines). The error bars represent the mean rms of the linear fits. The density maps show the two projections of galaxies where we can appreciate the existence of a ring system (upper panel) and a close encounter (lower panel).
Both simulated galaxies are at $z\sim 0.7$ and  the images correspond to  projected squares of 20 kpc a side. }
\label{gradientspositivos}
\end{figure*}

\begin{figure*}
\resizebox{4.3cm}{!}{\includegraphics{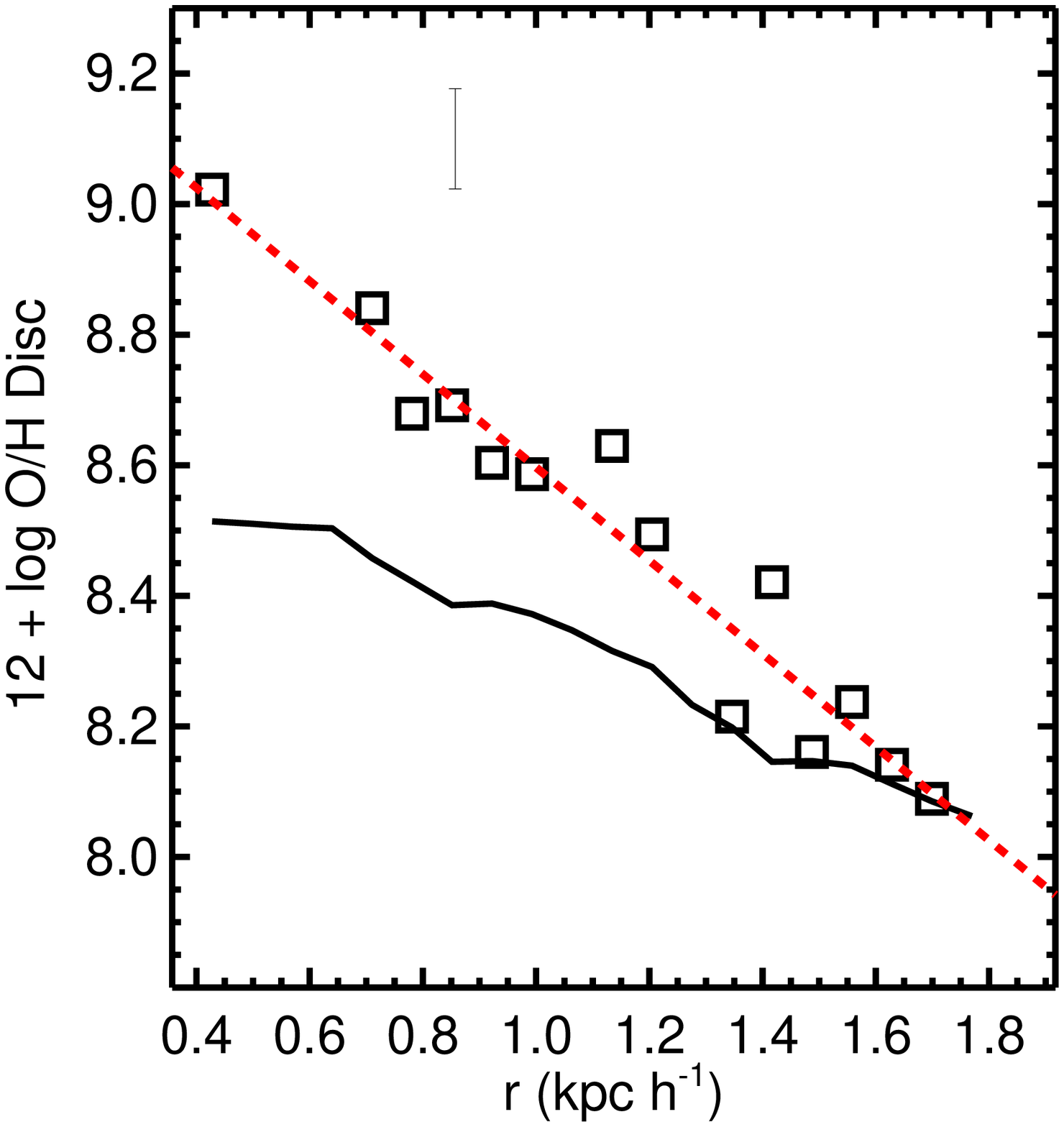}}
\resizebox{4.3cm}{!}{\includegraphics{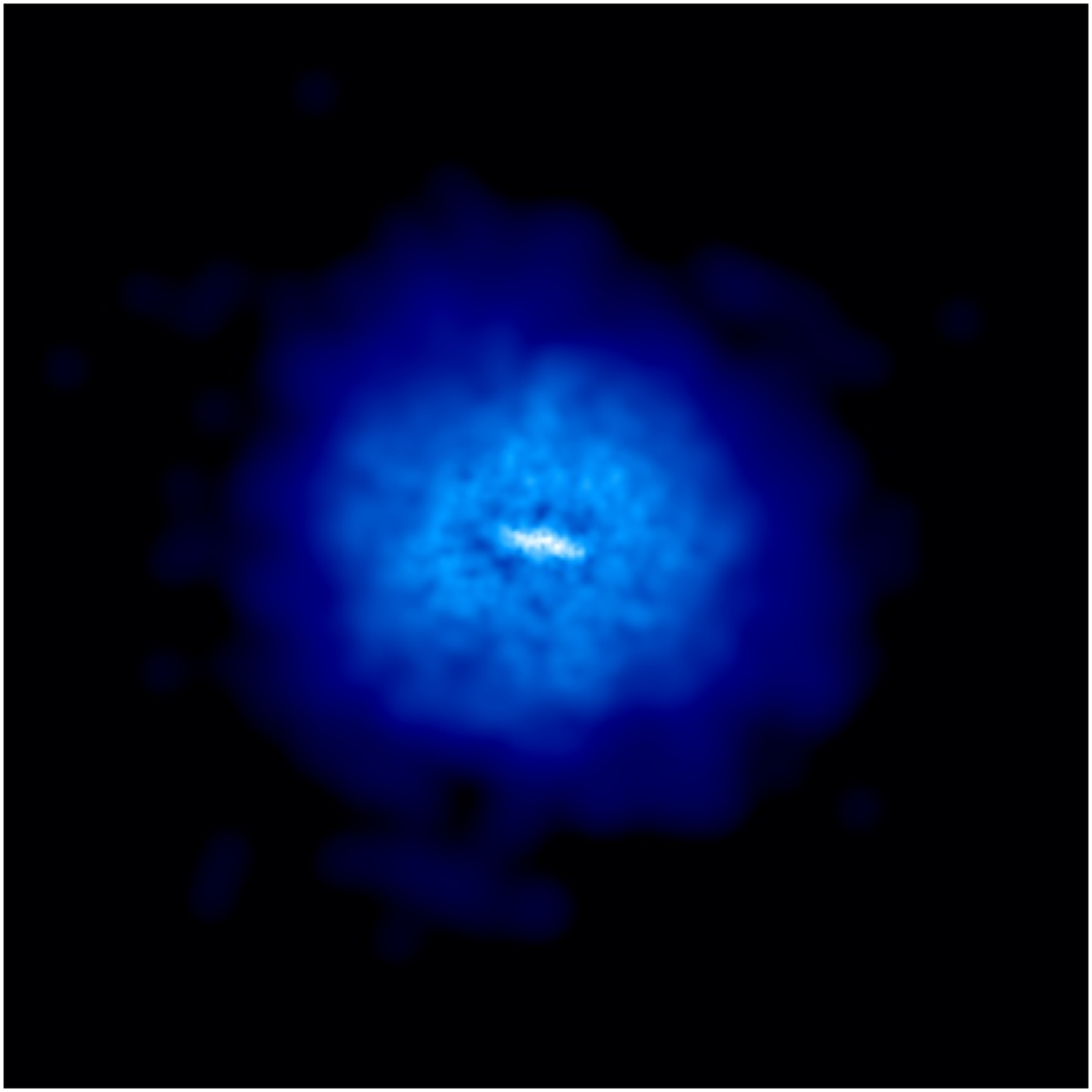}}
\resizebox{4.3cm}{!}{\includegraphics{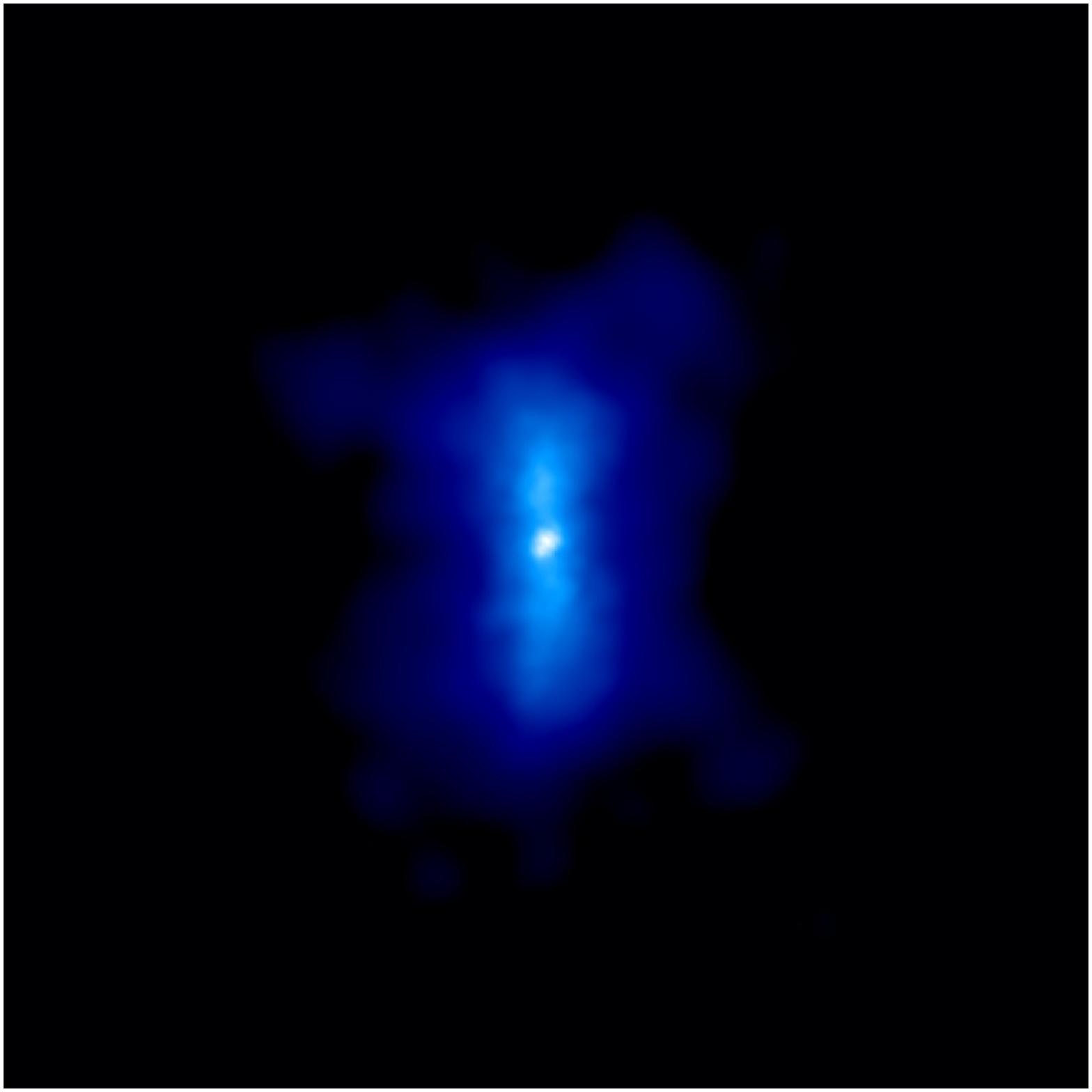}}
\resizebox{0.9cm}{!}{\includegraphics{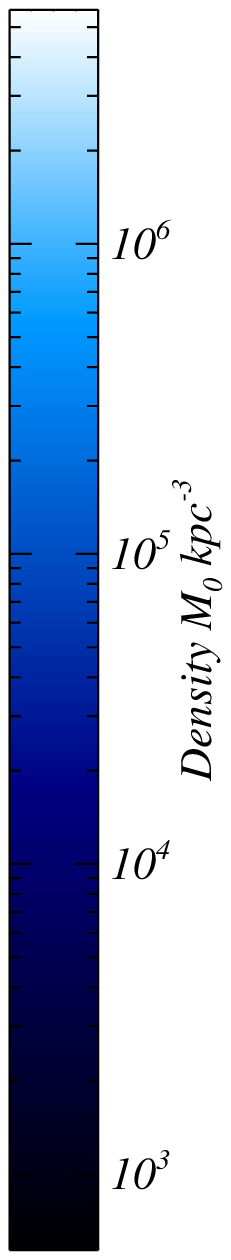}}\\
\resizebox{4.3cm}{!}{\includegraphics{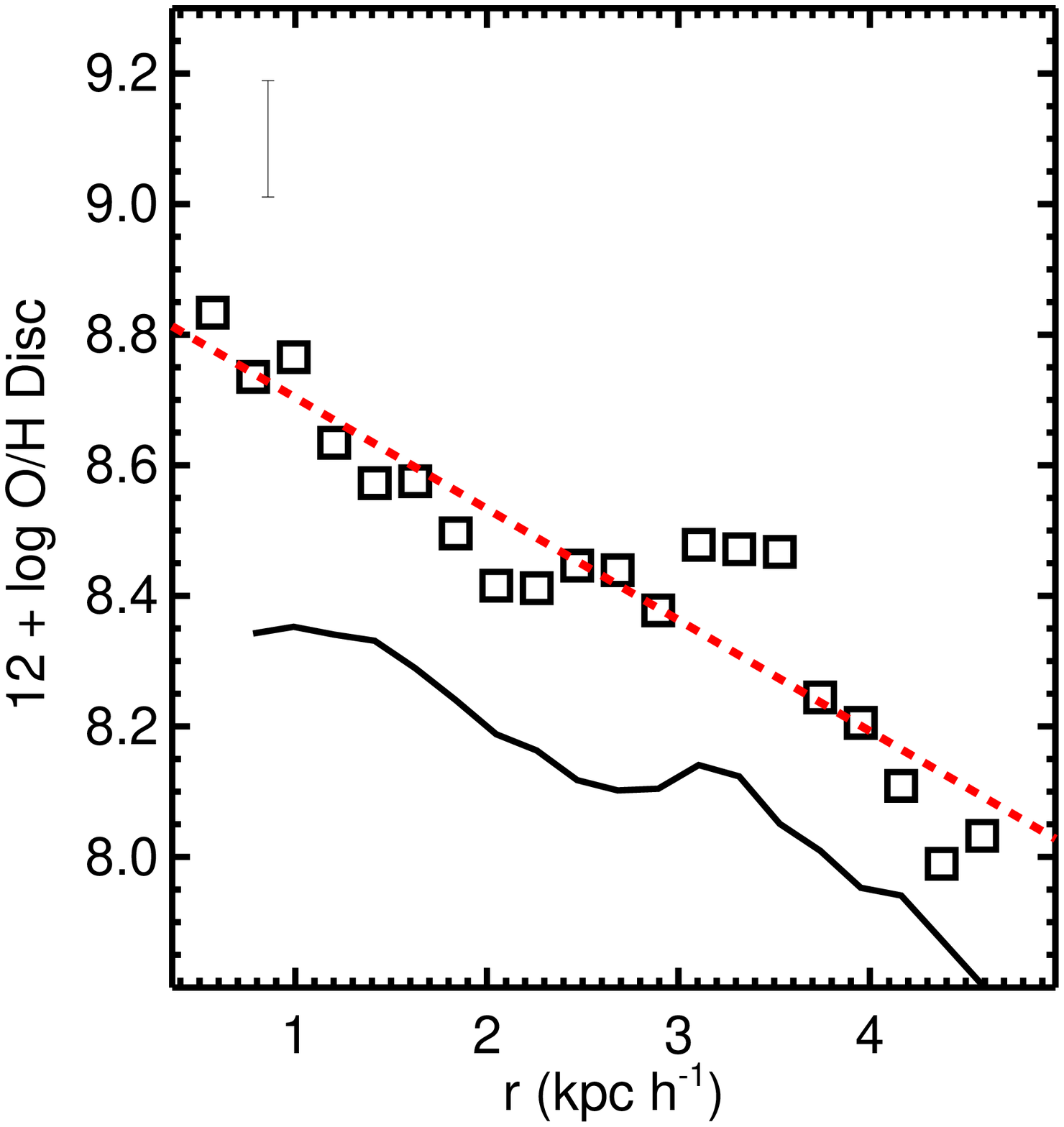}}
\resizebox{4.3cm}{!}{\includegraphics{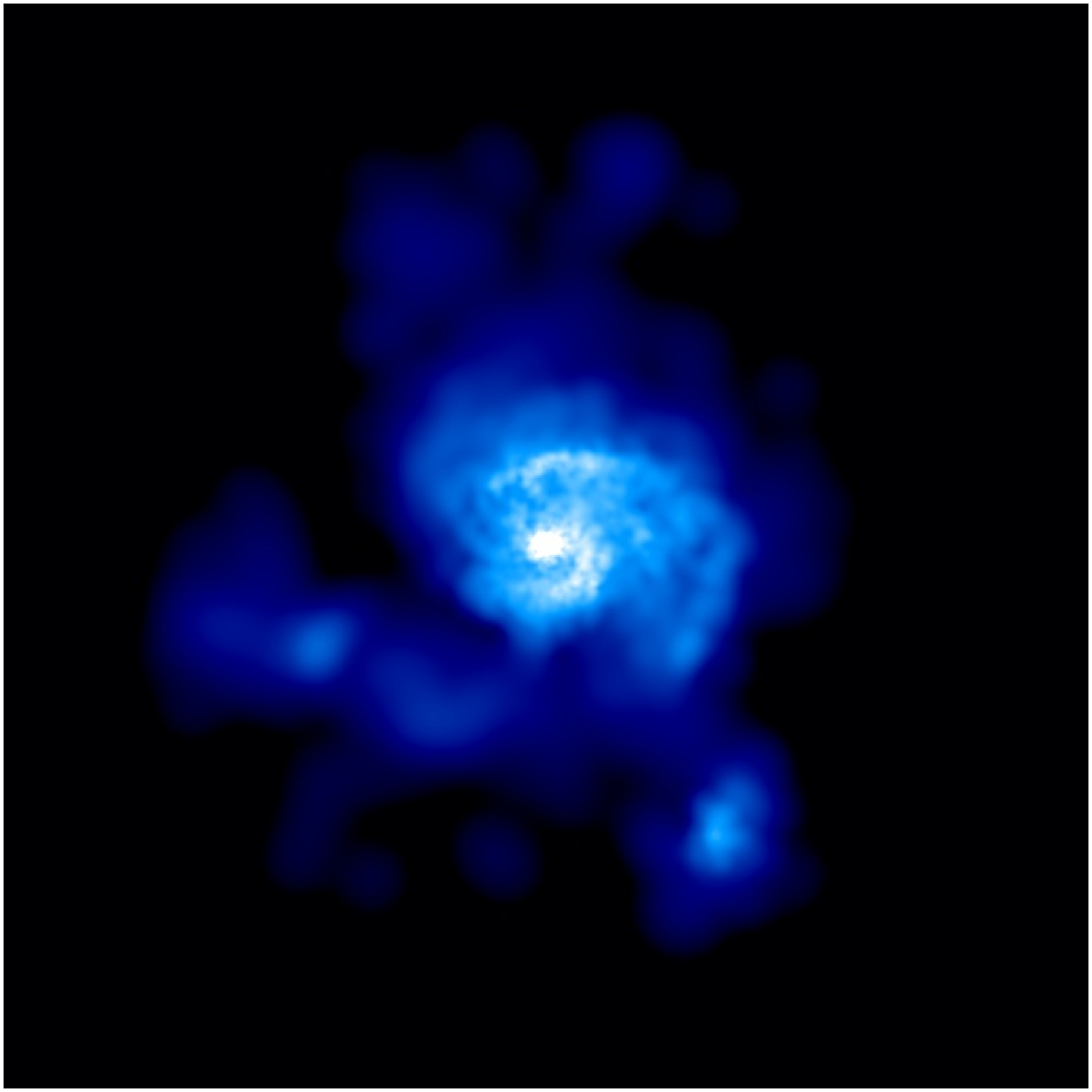}}
\resizebox{4.3cm}{!}{\includegraphics{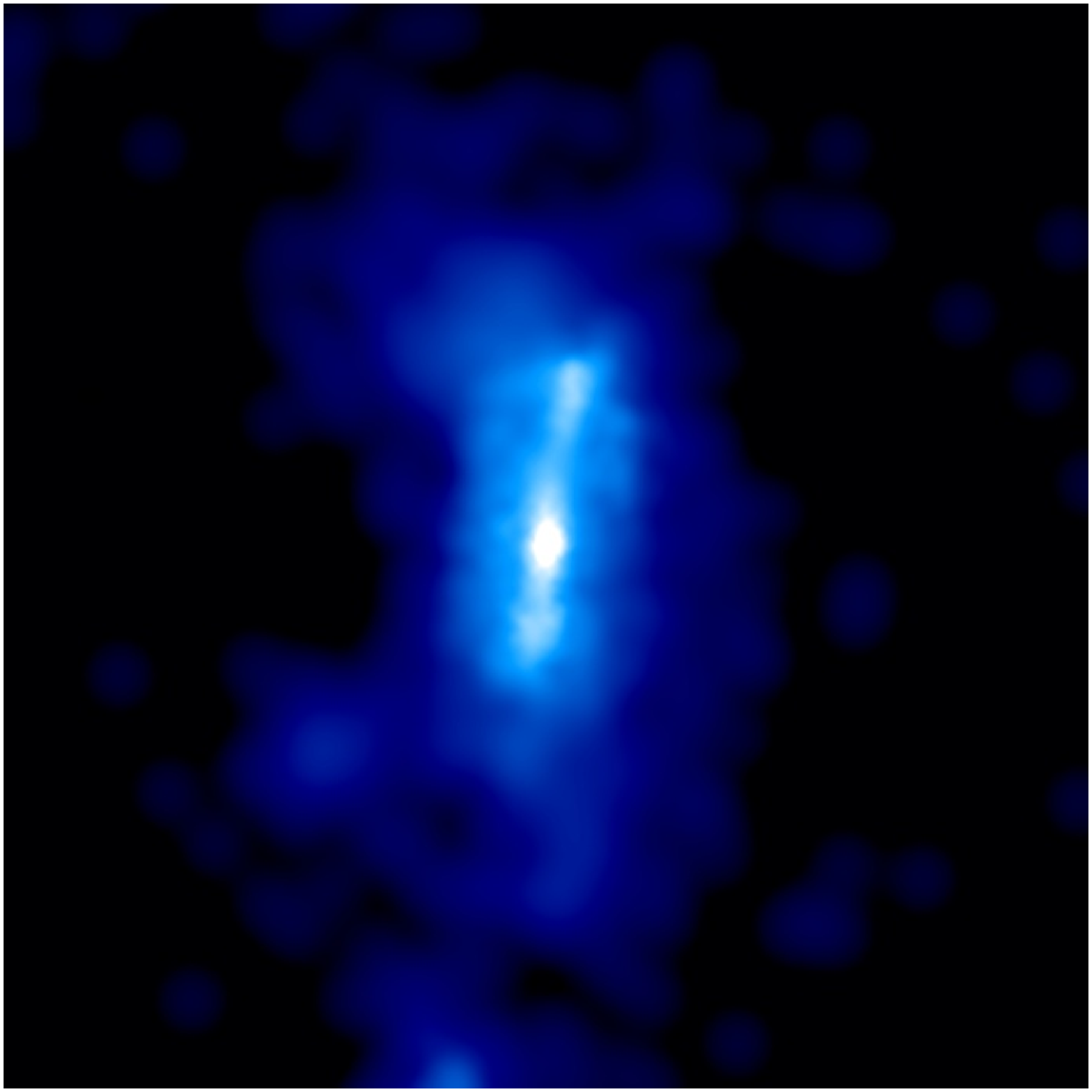}}
\resizebox{0.9cm}{!}{\includegraphics{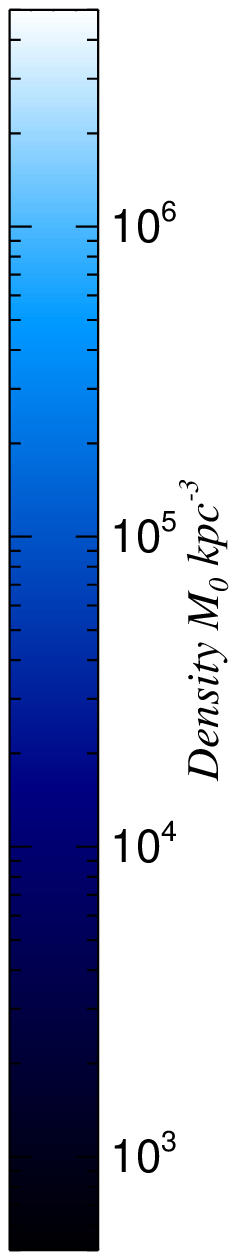}}
\caption{  Mean $12+$log~O/H profiles  determined for  the gas-phase disc components (open squares) and the  stellar discs populations (black solid curves) in  simulated  galaxies which show very negative
gas abundance profiles. 
The linear regressions to the gas abundance profiles are also included (red lines). The error bars represent the mean rms of the linear fits. The density maps show the two projections of galaxies where we can appreciate the existence of a concentrated bar (upper panel) and  a close interaction (lower panel). Both galaxies
are at $z \sim 2$ and  the images correspond to  projected squares of 10 kpc a side. }
\label{gradientsnegativos}
\end{figure*}

We found that gaseous discs with positive abundance gradients tend to exhibit morphological perturbations such as the presence of important central bars, 
clear ring structures in the discs or a very close companion, as can be seen from Fig.~\ref{gradientspositivos} where two examples are shown for illustration purposes.
Regarding discs with very negative slopes, we  find a more complex situation: some of them have close companions, others have central bars while the rest are consistent with having very concentrated mass distributions. Most of them
shows either indications of low-metallicity accretion as the gas components have lower or similar median chemical abundances than the stars at a given radius, or of recent strong star formation activity in the central regions as the median gas abundances are much higher than those of the 
underlying stellar populations. In  Fig.~\ref{gradientsnegativos} we show two examples.  We stress the fact  that this is a qualitative description since there are not enough time-steps to follow the mergers and interactions in our cosmological run.

In order to shed light on this point,   we resort to  the analysis  of the chemical evolution  of galaxy pairs. These experiments 
showed that mergers and interactions could indeed trigger gas inflows which transport low-metallicity material into
the central regions \citep{perez2006,dimatteo2009,rich2012}. In particular, \citet{perez2011} followed the evolution of the metallicity gradients along the interactions using hydrodynamical simulations including chemical evolution. These authors reported the triggering of low-metallicity inflows which contribute to flatten the inner abundance profiles. 
These authors also showed clearly  that
as soon as this material reaches the central regions, the star formation conditions are generally satisfied and {the gas} is promptly transformed into stars.  The new-born stars generate SNII which inject new $\alpha$-elements into the ISM, increasing the $\alpha$-abundances in the central regions and producing steeper negative oxygen gradients. These abundance profiles could become more positive again as galactic outflows
from the inner regions transport enriched material outside the galaxy. Even more this enriched material could contribute to the outskirts of the discs {if it cools down and is re-accreted by the main galaxy} \citep[see figure 7 in][]{perez2011}.  Hence, interactions could produce almost flat, positive slopes as well as  negative ones, depending on the stage of the
interactions, the gas availability and the characteristics of the disc systems \citep[see figures 6 and 11 in][]{perez2011}. 
Also the strength of the SN feedback could modulate the efficiency of these processes \citep[e.g.][]{pilkington2012}.

\subsubsection{The effects of mergers}

To explore the evolution of the metallicity gradient and the sSFR during  a galaxy-galaxy interaction, we used   the gas-rich major merger simulated by \citet{perez2011}. We choose this merger event since it provides
a strong reaction to the effects of tidal fields during the interactions due to the high-gas content of the interacting systems.
 The simulation follows the evolution of a 1:1 major merger event with a gas fraction of $50\%$ in the disc\footnote {The encounter is co-planar and the galaxies are co-rotating. Each galaxy is resolved with 200000 dark matter particles and 100000 initial gas particles with a mass of $\sim 3 \times 10^5$M$_{\odot}$. The gravitational softenings adopted are 0.16 kpc for the gas particles and 0.32 kpc for dark matter.  For more details on the initial conditions  see  \citet{perez2011}.}. An important aspect to take into account is that the initial metallicity profiles of the interacting galaxies have been set {\it ad hoc} to have a slope of ${-0.07~\rm dex~kpc^{-1}}$ and a mean chemical
abundance consistent with the mass-metallicity relation at $z\sim 2$ \citep{maiolino2008}. Hence, this initial oxygen slope should be taken 
only as indicative. The relevant aspect to consider here is the change of the slope of the abundance profiles  as a function of the sSFR,  along the interaction.

\begin{figure*}
\resizebox{14cm}{!}{\includegraphics{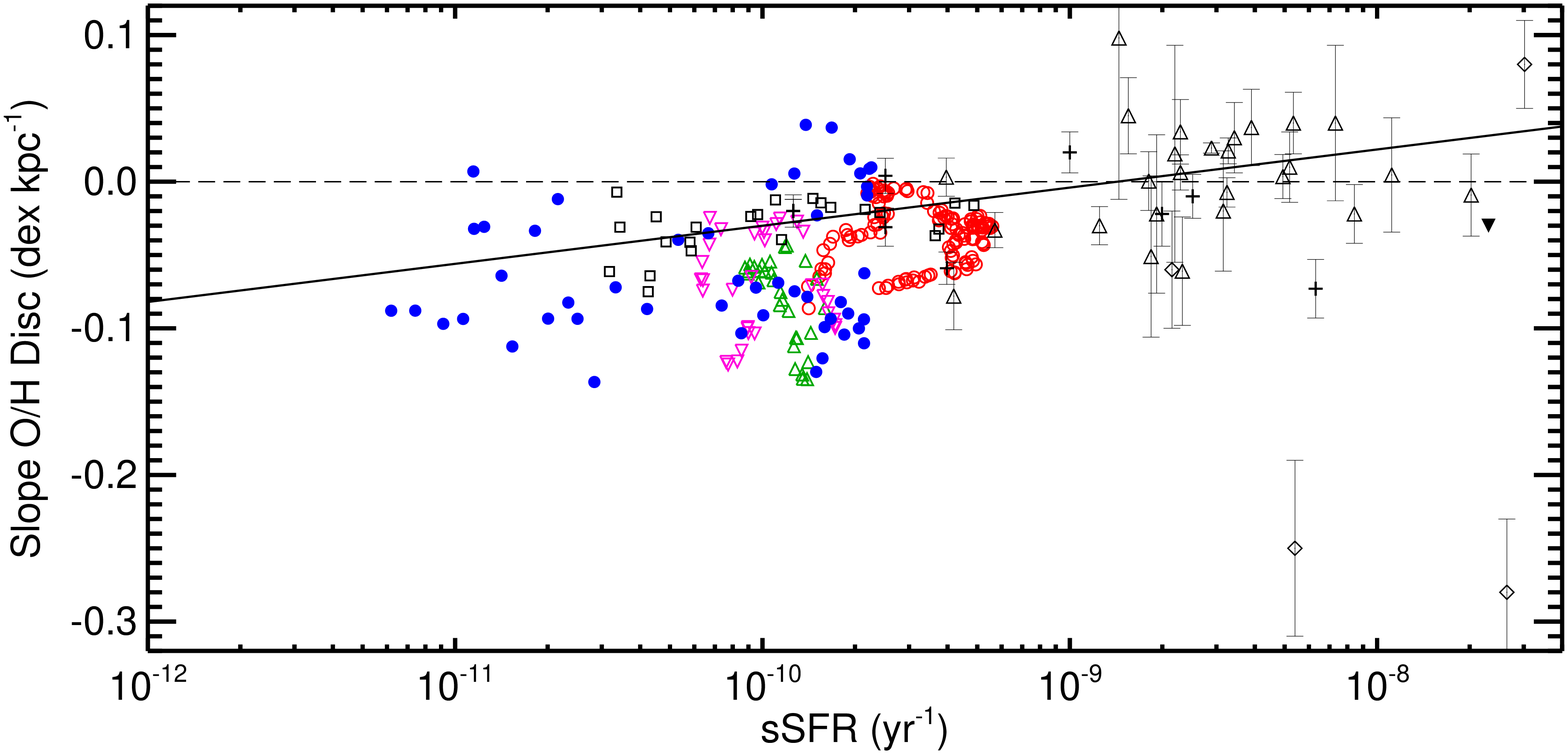}}
\hspace*{-0.2cm}
\caption{ Slope of the gas-phase oxygen profiles as a function of the sSFR for a galaxy  in a gas-rich major merger.
We denote the periods before the pericentre (red open circles), between the pericentre and apocentre (green triangles), between   the  apocentre and the second pericentre (magenta inverted triangles) and after the second pericentre (blue filled circles).
 For comparison we include the observational results from  \citet[black, open squares]{rupke2010}, \citet[black, open triangles]{queyrel2012}, \citet[black, open rombus]{jones2013} \citet[black, inverted triangles]{jones2015}, \citet[black, crosses]{stott2014}  and the linear regression  reported by \citet[solid line]{stott2014}.
The galaxy starts with an {\it ad hoc} metallicity slope of ${-0.07~\rm dex~kpc^{-1}}$. See  Fig.~\ref{slopessfr_merger2} and Section 3.2.1 for details. 
}
\label{slopessfr_merger}
\end{figure*}

\begin{figure*}
\resizebox{5cm}{!}{\includegraphics{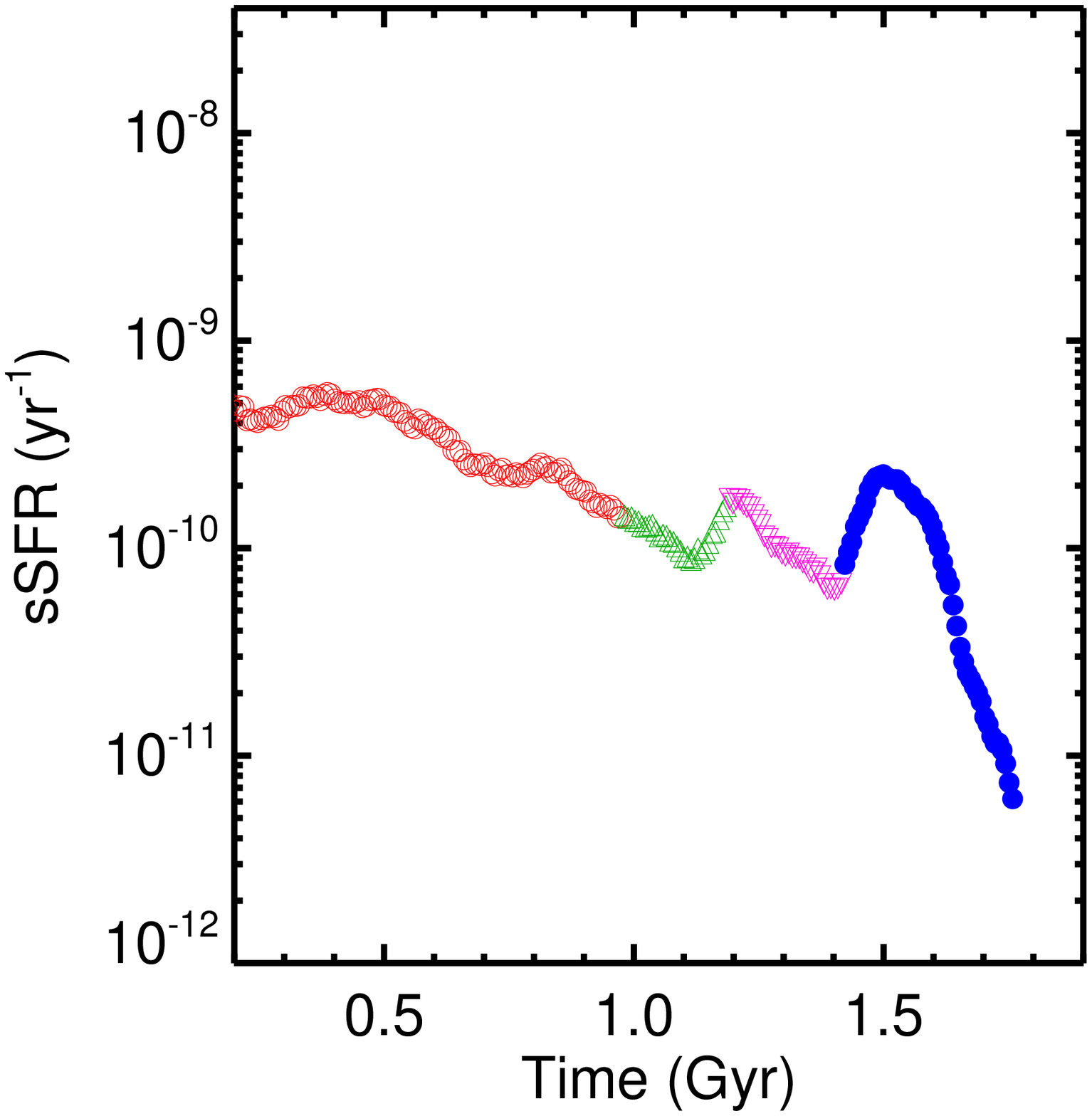}}
\resizebox{5cm}{!}{\includegraphics{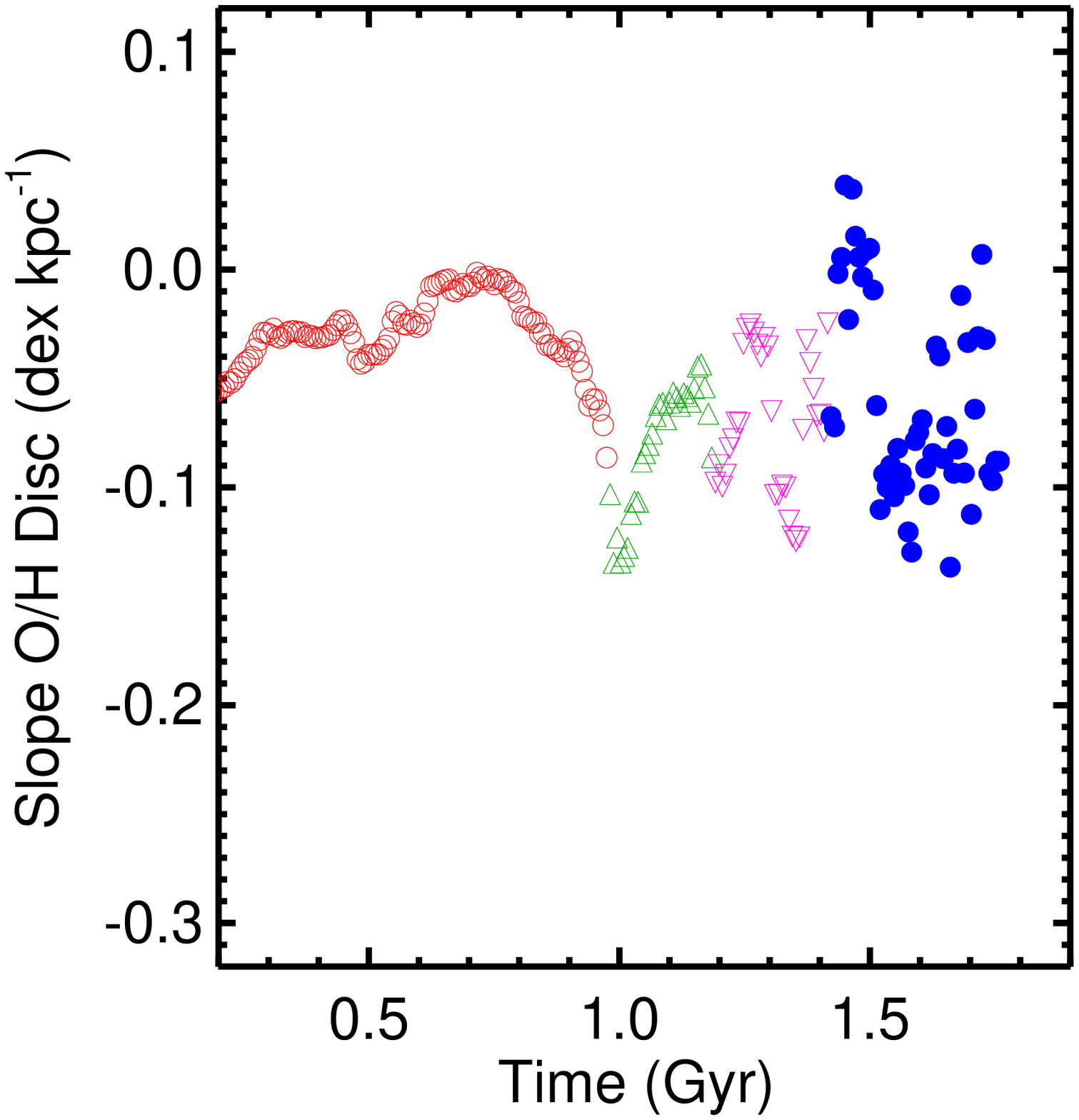}}
\resizebox{5cm}{!}{\includegraphics{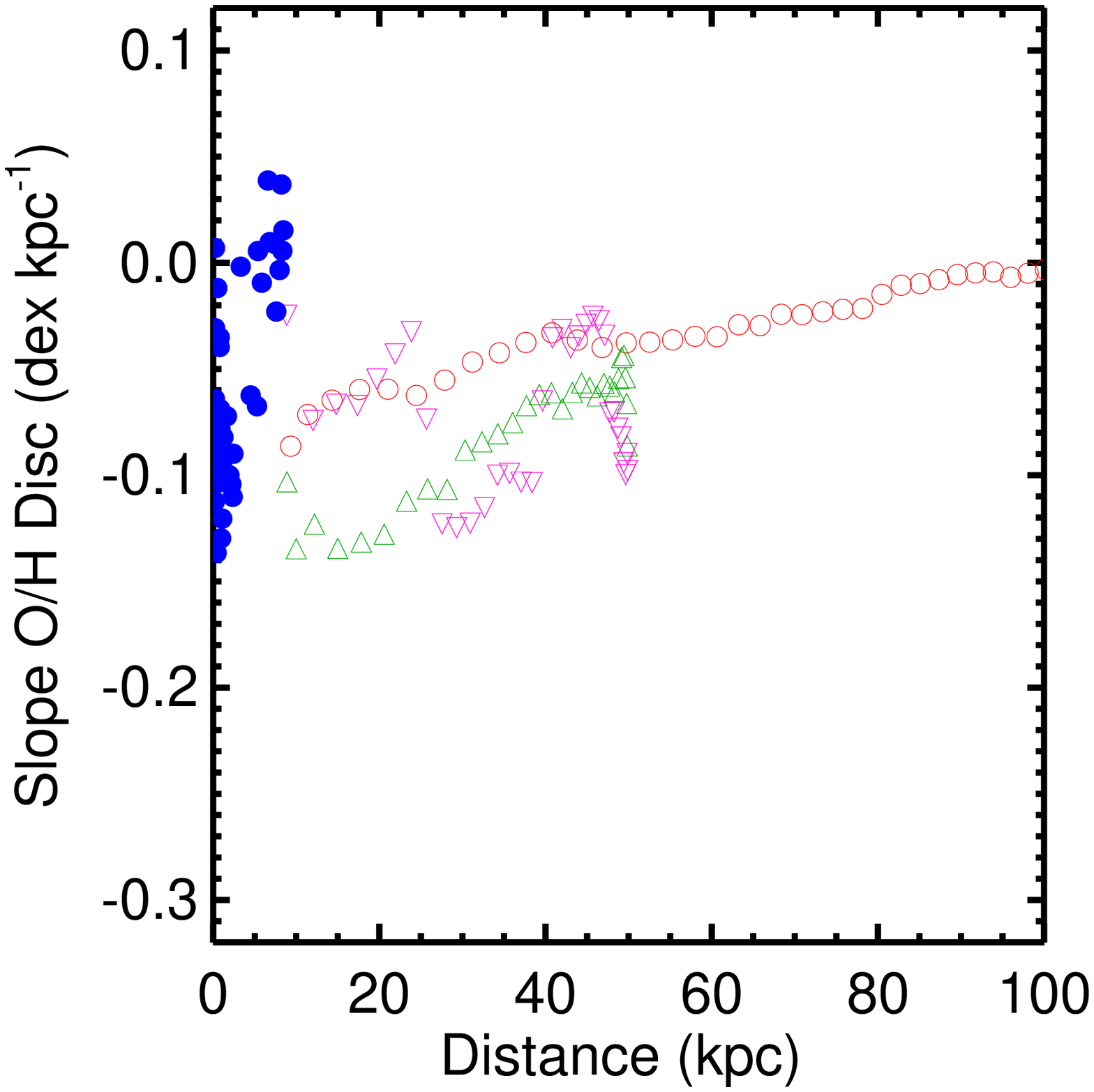}}
\hspace*{-0.2cm}
\caption{ Time evolution of sSFR (left panel) and the gas-phase oxygen gradients (middle panel). The evolution of the gradients as 
a function of distance between the mass centres are also included (right panel). Key periods of the encounter are shown in different colours (see Fig.~\ref{slopessfr_merger} for colour-code).}
\label{slopessfr_merger2}
\end{figure*}

In Fig.~\ref{slopessfr_merger}, we plot the sSFR versus the slope of the gas-phase oxygen profiles for the disc component in  one of the interacting galaxies.  
These parameters evolve along the encounter in a complex way. To  understand this evolutionary path, 
in Fig.~\ref{slopessfr_merger2} we show the metallicity gradient as a function of the relative distance between 
the mass centre and as a function of time. 
The time evolution of the  sSFR is also included. 

At the beginning the sSFR is high  because of the high gas fraction in the galaxies and the slope of the metallicity gradient is  close to $-0.07{~\rm dex~kpc^{-1}}$ as expected. 
As the systems approach each other,  the slope of the metallicity gradients  becomes less negative as 
  gas inflows are driven inwards, 
carrying  low-metallicity material \citep[see also figures 8, 9 and 10  in][]{perez2011}. The increase
of the gas  density produces new stars and hence, new chemical elements are injected into the ISM increasing the oxygen abundances. As a consequence, 
 the slope of the metallicity gradient becomes negative again as can be seen from Fig.~\ref{slopessfr_merger2} (middle panel).
 After the apocentre and  as galaxies approach each other again,  positive slopes are detected. 
By this time, the galaxies are very close (less than $\sim 10$ kpc).  Negative slopes are determined by the 
subsequent enrichment produced by the new stars. From this figure we can see that the behaviour is not smooth but the evolution
of the metallicity gradients as a function of time shows  a series of saw tooth as well as does the sSFR.
These plots clearly illustrate how the increase of the sSFR can
 be related to the flattening of the metallicity gradients and how they  become more negative as new stars inject 
massively $\alpha$-elements in the central regions. This can also be correlated with an increase of the [O/Fe] ratio as
reported by \citet[][figure 8d]{perez2011}.

If galaxies formed consistently within a $\Lambda$-CDM cosmology, mergers and interactions  would be ubiquitous, mainly in the  high-redshift Universe. 
These violent events could cause different perturbations in the metallicity gradients, providing  a range of situations which are not covered
by the equal-mass, gas-rich  merger  investigated in this Section. However, this simulation illustrates clearly a possible evolutionary  path for 
the metallicity gradients during
interactions and accounts for the negative and the positive slopes found in the discs galaxies simulated  in a cosmological framework.

\section{Conclusions}

We studied the chemical abundance profiles
of the  gas-phase  disc components in relation to the star formation activity of the galaxies simulated in a hierarchical universe.  
For this purpose, we analysed a set of galaxies extracted from a cosmological  simulation consistent with a $\Lambda$CDM scenario, performed using  a hydrodynamical code which
includes energy and chemical feedback by SNII and SNIa. 
We performed the analysis as a function of stellar mass in the range $[10^9-10^{11}]$ M$\odot$ by adopting a stellar-mass limit of $M_{\rm star} = 10^{10}$M$\odot$. 
The  global SFR of the simulated galaxies as a function of stellar mass are consistent with observational values.

Our main results can be summarized as follows:
\begin{itemize}
\item At $z\sim 0 $ the   gas-phase oxygen slopes of the simulated discs show  a trend with the stellar mass in agreement with observations. 
Galaxies with lower stellar masses have  metallicity gradients which exhibit larger dispersion than those measured in galaxies with larger masses, in agreement with observations results \citep{ho2015}. 

Massive galaxies show very weak evolution of their oxygen gradients with redshift (i.e. the change in the mean oxygen slopes between $z\sim 0$ and $z\sim 2$ is within the bootstrap dispersion).
Conversely,  galaxies with lower stellar masses show more significant changes of the metallicity gradients with redshift, so that they get steeper with increasing redshift. At $z\sim 1$ our  simulated discs exhibit slighlty steeper
negative metallicity slopes than those reported by current  observations.   At $z \sim 2$ there are few available observations which have diverse metallicity gradients.
 At these high redshifts, the simulated metallicity slopes  also  show  a noticeable scatter for galaxies with low stellar masses, but the mean value is dominated by
the  negative gradients.

The fraction of gaseous discs with slopes smaller than $ -0.1~{\rm dex~kpc^{-1}}$  in galaxies with  low stellar-masses varies from 0.10  at $z\sim 0$ to 0.78  $z\sim 2$. In the high stellar-mass subsample the fraction is small or null. Conversely, the fraction of disc galaxies with positive slopes is found to be approximately constant for galaxies  with low stellar masses ($\sim 0.10$ on average) but to increase with increasing redshift for 
the galaxies with higher stellar masses  (from $\sim 0.05$ at $z\sim 0$  to $\sim 0.17$ at  $z \sim 2$).

\item  At $z\sim 0$, galaxies with  both low and high stellar masses have mean metallicity slopes and sSFR  consistent with the relation reported by \citet{stott2014}. 
Galaxies with high stellar masses have values which are consistent with  the Stott's relation in the three redshift interval analysed. However, galaxies with the lower masses deviate from the
observed relation for  higher redshift as the results  of the increasing  contribution of  gaseous discs with metallicity slopes smaller than $-0.1~{\rm dex~kpc^{-1}}$.
If disc galaxies with such negative slopes are not included, then the mean values for both low and high stellar-mass galaxies are consistent with observed relations. We note  that \citet{stott2014} did not include the steeper negative slopes reported by \citet{jones2013}. 
However, in this case the correlation between {the slope of the oxygen gradients} and stellar mass is lost. At $z\sim 0$, 
this is in open disagreement with observational results, suggesting that a fraction of steep negative slopes should be present for low stellar-mass galaxies \citep{ho2015}.
At higher redshift,  a large observational data sample with more precise metallicity estimations is  required in order to study the presence and frequency of  very negative  and positive slopes as a function of stellar mass. 

\item  We explore the morphology and surrounding media of our simulated galaxies finding that  galaxies with positive and very negative metallicity slopes  have disturbed morphologies and close neighbours. 
 However, due to the low-number sample, it is not possible to derive a stastically robust trend. However, since they all have a system of surrounding satellites (of different masses) the probability of interactions/mergers
is expected to be high. The analysis of a simulated merger of two  gas-rich equal-mass  galaxies  yields that both negative and positive metallicity gradients might be produced during different stages of evolution of the encounter. 
It remains to be investigated if internal instabilities (such as secular evolution)
 could also lead to  a similar relation between the sSFR and the metallicity gradients and if the predicted relation can be distinguish from that of galaxies with
mergers and interactions.
 
\end{itemize}

Our results provide an interpretation to the observed relation reported by \citet{stott2014}. However,  a significant number of  negative slopes,  such as those reported by \citet{jones2013},  are detected for simulated galaxies with  low stellar masses. The presence of  such metallicity profiles needs to be confirmed by  larger observed galaxy samples.
In such case the location of a disc galaxy on this relation might tell us about its current evolutionary status as well as its past history.
Otherwise, it might suggest that our simulations are  overpredicted them, and hence our subgrid physics should be accordingly revised. As shown in previous works, the characteristics of the subgrid physics have an important impact on the
chemodynamical properties of galaxies \citep[e.g.][]{pilkington2012,gibson2013,snaith2013}. Hence the confrontation of the chemical properties of galaxies with observation is indeed a powerful tool to
set constrains on  galaxy formation models.

\section*{Acknowledgments}
{We thank the anonymous referee for the careful reading of the paper and constructive suggestions. We acknowledge J. Stott and  P.Norberg for useful comments. Simulations are part of the Fenix Project and  have been run in  Hal Cluster of the Universidad Nacional
de C\'ordoba, AlphaCrucis  of
IAG-USP (Brasil) and Barcelona Supercomputer Center. We acknowledged the use of   Fenix Cluster of
Institute for Astronomy and Space Physics.
This work has been partially supported by PICT
Raices 2011/959 of Ministry of Science (Argentina), Proyecto
Interno UNAB,  Project Fondecyt 2015 Regular and Nucleo UNAB (Chile). JMV acknowledges funding by project AYA2013-47742-04-01 of the Spanish PNAYA.
}

\bibliographystyle{mn2e}

\def\apj{ApJ}
\def\apjl{ApJ}
\def\aj{AJ}
\def\mnras{MNRAS}
\def\aa{A\&A}
\def\nat{Nature}
\def\araa{ARA\&A}
\def\aap{A\&A}

\bibliography{Tissera_rv3}

\IfFileExists{\jobname.bbl}{}
{\typeout{}
\typeout{****************************************************}
\typeout{****************************************************}
\typeout{** Please run "bibtex \jobname" to optain}
\typeout{** the bibliography and then re-run LaTeX}
\typeout{** twice to fix the references!}
\typeout{****************************************************}
\typeout{****************************************************}
\typeout{}
}

\bsp	
\label{lastpage}
\end{document}